\documentclass[11pt,openany,oneside]{article}
\usepackage[papersize={210mm,297mm},hdivide={20mm,*,20mm},vdivide={25mm,*,25mm},dvips,mag=1000]{geometry}

\def\bSig\mathbf{\Sigma}

\usepackage[T1]{fontenc}
\usepackage{parskip}
\usepackage{latexsym,amsmath,amssymb, bm, caption, subcaption, bbding, amsthm}
\usepackage{graphicx}
\usepackage{times}
\usepackage{zi4}
\usepackage[toc,page]{appendix}
\usepackage{multirow, multicol}
\usepackage{xcolor, colortbl}
\usepackage{tabularray}
\usepackage{bbm}
\usepackage{pdfpages}
\usepackage{float}
\usepackage{epstopdf}
\usepackage{longtable}
\usepackage{algorithm, algpseudocode}
\usepackage{soul}
\usepackage{changepage}
\usepackage{natbib}
\usepackage{hyperref}
\hypersetup{colorlinks,allcolors=black}
\usepackage[figuresright]{rotating}

\newtheorem{theorem}{Theorem}
\newtheorem{lemma}[theorem]{Lemma}
\newtheorem{proposition}[theorem]{Proposition}

\newtheorem{definition}{Definition}
\theoremstyle{definition}

\newcommand\summaryname{Abstract}
\newenvironment{Abstract}%
    {\small\begin{center}%
    \bfseries{\summaryname} \end{center}}

\begin{document}
\setlength{\parindent}{0mm}
\begin{center}
\Large
\newfont{\cmd}{cmdunh10 scaled \magstep3}
\vspace*{-20mm} \hspace*{-0,5cm}
\thispagestyle{empty} 

\vspace{1cm}  
{\LARGE\textbf{Design specification of Partial Ordering Continual Reassessment Method based on consistency conditions}}

Weishi Chen$^1$ and Pavel Mozgunov$^2$\\
$^1$ \textit{weishi.chen@mrc-bsu.cam.ac.uk}\\
$^2$ \textit{pavel.mozgunov@mrc-bsu.cam.ac.uk}\\
$^{1,2} $\textit{MRC Biostatistics Unit, University of Cambridge, Cambridge, UK}
\vspace{1cm}
\end{center}

\begin{Abstract}
\begin{changemargin}{1cm}{1cm}
The study of combinations of drugs/drug-schedules gained increasing attention in various therapeutic areas recently. In oncology, the aim of phase I combination clinical trial is to find the maximum tolerated combination (MTC). Many innovative designs were proposed, among which the Partial Ordering Continual Reassessment Method (POCRM) is increasingly applied due to its simplicity and versatility. The POCRM requires specification of plausible monotonic orderings of combinations. However, the choice remains a major difficulty, especially in trials with many compounds or/and combinations. Practical recommendations are given to select six orderings based on statistical considerations, while simulation studies found the design performs poorly when the MTC is in the middle of the combination grid. We prove that the POCRM under currently recommended orderings can be inconsistent (i.e., cannot achieve 100\% correct selection even under infinite samples) which translates into poor performance under small sample size. Based on the derived consistency conditions, we provide two practical recommendations on how to select orderings for real studies (i) based on plausible combination-toxicity scenarios, (ii) regardless of the possible scenarios. We also provide guidance on how to choose other design parameters based on the asymptotic properties and demonstrate how it improves small sample behaviours.
\end{changemargin}
\end{Abstract}

%




%

\section{Introduction}

Phase I clinical trials is the first instance new treatments are given to human, and the focus of these studies is on safety. In oncology for cytotoxic drugs, it is often reasonable to assume both efficacy and toxicity of the treatment increase with dose level, the monotonicity assumption. Therefore, one of the conventional aims of phase I is to find the highest dose having a toxicity probability within a tolerable level, i.e., the maximum tolerated dose (MTD). There is a recognised need for more efficient and data-driven approaches to identify the right dose in terms of the toxicity. At the same time, there is also an increasing interest in methods trying to find an ``optimal'' dose when the higher toxicity does not necessarily translate into a higher efficacy (e.g. for cytotoxic drugs). Project ``Optimus'' initiated by FDA~\citep{FDA} seeks to challenge how dose-finding is currently undertaken, specifically, by taking into account non-toxicity outcomes when exploring different doses. However, many Phase I oncology trials of new experimental agents are conducted in all-comers and no immediate clinical benefit might be observed for those patients (e.g. immunotherapies like inhibitors). As a result, the safety component of these trials remains the paramount part of Phase~I drug development to safeguard the patients and, hence, adequate safety modelling remains a core part of Phase I studies and its efficiency is of paramount importance.

A number of phase I designs of single-agent dose-escalation trials have been proposed, ranging from rule-based~\citep{Storer1989}, model-assisted~\citep{Liu2015BOIN}, to model-based \citep{OQuigley1990CRM} and their novel extensions. The first model-based design proposed was Continual Reassessment Method (CRM) proposed by \cite{OQuigley1990CRM}. The fundamental idea of model-based designs is to use a (simple) parametric model between dose levels and toxicity probabilities, and then estimate the model parameters based on data and clinical knowledge. It has been shown since the original proposal that the CRM has favourable operating characteristics and is most efficient in identifying the MTD~\citep{Paoletti2004DesignEfficiency, Hansen2014}. 

Furthermore, studies of dose-combinations are attracting increasing clinical interest due to their potential to achieve better efficacy results or better tolerability of the drugs. In cases where the doses of several agents can be changed, the aim is to find the \textit{maximum tolerated combinations} (MTC). The main challenge is to determine the toxicity ordering between combinations whose ordering cannot be implied by monotonicity on each single agent. Many of the single-agent dose-finding designs have been extended to combination trials, see \citet{Lin2017BoinCombinations}~for combination model-assisted design, \citet{Mozgunov2020SurfaceFree}~for model-free designs, and \citet{Riviere2014ModelBasedCombination}~for model-based designs. 

An extension of the CRM, Partial Ordering CRM (POCRM), to accommodate the uncertainty in the toxicity ordering between combinations was proposed by \cite{Wages2011TwoStageCRM}. Assuming several potential orderings of combinations (instead of one in the CRM), Bayesian model selection is applied to select the ordering most compatible with data under which, the CRM is applied to estimate the combination-toxicity relationship. It has been shown that the performance of the POCRM is superior among designs for combination trials~\citep{Hirakawa2015ComboComparison}. Due to its versatility, the POCRM is extended to other trial settings, such as dose-scheduling or the combination of more agents~\citep{Wages2014DoseSchedule}. Moreover, the type of outcomes can be extended from the original binary to other types, such as time-to-event~\citep{Wages2012TITEPOCRM} and toxicity/efficacy~\citep{Wages2015SeamlessI/II, Mozgunov2019Entropy}. Hence, the POCRM is increasingly often used in practice \citep{Mozgunov2022PracticalImplementation, Wages2024POCRMTrials, Yap2024ASCO}, see e.g. the real trials Mel 58, Breast 49, ABT-199 (\href{https://clinicaltrials.gov/}{ClinicalTrials.gov Identifier}: NCT01585350, NCT03473639, NCT02419560, respectively). 

While the POCRM can accommodate many trial settings, it requires several design parameters to be specified, most importantly, the potential monotonic orderings to be specified prior to the start of the trial. The number of orderings increases combinatorially with the number of combinations, and including all orderings can be computationally infeasible even with a moderate number of combinations. Practical recommendations have been given to include only six orderings based on statistical considerations, however, the performance can be poor if the true MTC is in the middle of the correct ordering~\citep{Wages2013Orders}. The difficulty in specifying orderings as well as this potentially compromised performance hinders the usage of the POCRM in practical applications, and therefore, it is important to have a more systematic way to specify orderings which has good performance under all possible scenarios of true toxicity probabilities.

We have found that the reason for the undermined performance under some combination-toxicity scenarios is that the POCRM might not be consistent with the specified set of orderings, i.e. does not select the correct MTC with probability one even as the sample size goes to infinity. Although phase I clinical trials typically have very small sample sizes, this inconsistent behaviour can also translate to poor operating characteristics in real trials as it hinges the fundamental inability of a design to select the correct combination. We will show that by exploring the asymptotic properties, a deeper understanding of the design can be obtained, based on which, ordering specification can be done in a systematic way that guarantees consistency, but also improves the performance at small sample sizes. 

Indeed, the utility of understanding the theoretical properties of the design has proven to be useful for the practical application of the design. \cite{Shen1996Consistency}~provided sufficient conditions for consistency of the CRM, which were relaxed by~\cite{Cheung2002RelaxedConsistency}, and subsequently extended to two-parameter models by~\cite{OQuigley2006Theoretical}. These conditions provide an insight into the performance and mechanisms of the CRM design together with important directions on how to set up some design parameters of the CRM. For example, it is used as a foundation of the indifference interval technique, which provides well-calibrated skeletons under a wide range of scenarios~\citep{LeeCheung2009Skeleton}. Besides, it serves as the central part of the simulation-free method to estimate the proportion of correct selection (PCS) of the MTD, which greatly saves computation~\citep{Braun2020SimFreeCalibration}. 

Little theoretical studies are known for any dose-combination design, including the POCRM, mainly due to the challenges of unknown ordering. In this work, sufficient conditions for the consistency of POCRM are derived. Based on these, we provide practical guidance on its ordering specification. Specifically, we provide two recommendations that are either (i) specific to given settings or (ii) agnostic to the combination-toxicity scenarios. In addition, we use the consistency conditions to guide the calibration of the design parameters to achieve better operating characteristics for small sample sizes.

\section{Motivating setting}
\label{Subsec: motivation}
The example below is motivated by a real trial, to which the authors contributed. Consider the $3\times3$ combination case, where drug A and B each has 3 levels. Let $\tilde{d}_{i,j}$ denote the combination $(a_i,b_j)$, $i,j=1,2,3$, which can be arranged into a matrix as in Table~\ref{Tab: 3x3 grid}.
\begin{table}[H]
\centering
\begin{tabular}{c c c c}\hline
    & \multicolumn{3}{c}{Drug A}\\ \cline{2-4}
    & $\tilde{d}_{1,3}$=($a_1,b_3$) & $\tilde{d}_{2,3}$=($a_2,b_3$) & $\tilde{d}_{3,3}$=($a_3,b_3$)\\
    \multirow{-2}{*}{Drug} & $\tilde{d}_{1,2}$=($a_1,b_2$) & $\tilde{d}_{2,2}$=($a_2,b_2$) & $\tilde{d}_{3,2}$=($a_3,b_2$)\\
    \multirow{-2}{*}{B} & $\tilde{d}_{1,1}$=($a_1,b_1$) & $\tilde{d}_{2,1}$=($a_2,b_1$) & $\tilde{d}_{3,1}$=($a_3,b_1$)\\ \hline
\end{tabular}
\caption{$3\times3$ dose-combinations.\label{Tab: 3x3 grid}}
\end{table}
Monotonicity of each single drug implies \textit{partial orderings} of dose-combinations. The term ``partial'' is used for subsets of the 9 combinations with increasing toxicities. Whereas, the term ``complete" is for whole 9 combinations ordered respecting the known partial orderings. For example, when fixing the level of B at $b_1$, monotonicity on drug A implies partial ordering $\tilde{d}_{1,1}\to\tilde{d}_{2,1}\to\tilde{d}_{3,1}$. Then, fixing A at $a_3$, monotonicity on B further implies partial ordering $\tilde{d}_{3,1}\to\tilde{d}_{3,2}\to\tilde{d}_{3,3}$. However, the ordering between $\tilde{d}_{1,2}$ and $\tilde{d}_{2,1}$ remains unknown, as $\tilde{d}_{2,1}$ has a higher level of A but lower level of B than $\tilde{d}_{1,2}$. When the number of combinations gets large, the number of possible complete orderings grows combinatorially. In this $3\times3$ case, there are 42 complete orderings (listed in Section~2 in the Supplementary Materials). It is computationally expensive to include all orderings and might be infeasible to communicate the meaning of all 42 to the clinical team. To tackle this, one can follow the suggestion by~\cite{Wages2013Orders} and include 6 orderings (detailed in Section~\ref{Subsec: number of models}). This will be referred to as ``Wages 6 orderings".

In the simulation below, the toxicity skeleton
$\bm{\alpha}^{(0)}$=(0.10, 0.20, 0.30, 0.40, 0.45, 0.50, 0.54, 0.59, 0.64) was used. The sample size is set to $n=60$. Table~\ref{Tab: motivation 3x3} below gives the proportion of correct selection (PCS) of the MTC, (based on $10^4$ simulations) under the POCRM with two choices of ordering specifications under scenarios 1-9 of toxicity probabilities defined in Table~\ref{Tab: scenarios}, where the target toxicity level (TTL) is $\theta_0=0.3$ and the MTCs are combinations with toxicities equal to 0.3. The prior probabilities of each ordering are equal. The orderings explored are (with the number corresponding to the ordering number as defined in Section~2 in the Supplementary Materials) (1) Wages 6 orderings: orderings 1, 14, 19, 23, 31, 35; (2) consistent 6 orderings (proposed): orderings 5, 11, 16, 19, 33, 42. 

\begin{table}[!htb]
    \centering
    \begin{tabular}{l ccc c ccc c ccc c ccc}
        \hline
        & \multicolumn{3}{c}{A} && \multicolumn{3}{c}{A} && \multicolumn{3}{c}{A} && \multicolumn{3}{c}{A}\\ \cline{2-4}\cline{6-8}\cline{10-12}\cline{14-16}
        \multirow{-2}{*}{B} & $a_1$ & $a_2$ & $a_3$ && $a_1$ & $a_2$ & $a_3$ && $a_1$ & $a_2$ & $a_3$ && $a_1$ & $a_2$ & $a_3$\\ \cline{2-16}
        & \multicolumn{3}{c}{Scenario 1} && \multicolumn{3}{c}{Scenario 2} && \multicolumn{3}{c}{Scenario 3} && \multicolumn{3}{c}{Scenario 4}\\
        $b_3$& 0.60 & 0.65 & 0.70 && 0.40 & 0.60 & 0.65 && 0.40 & 0.45 & 0.50 && 0.45 & 0.55 & 0.60\\
        $b_2$& 0.45 & 0.50 & 0.55 && 0.35 & 0.45 & 0.55 && 0.15 & 0.25 & 0.35 && \textbf{0.30} & 0.40 & 0.50\\
        $b_1$& \textbf{0.30} & 0.35 & 0.40 && 0.25 & \textbf{0.30} & 0.50 && 0.10 & 0.20 & \textbf{0.30} && 0.20 & 0.25 & 0.35\\\hline

        & \multicolumn{3}{c}{Scenario 5} && \multicolumn{3}{c}{Scenario 6} && \multicolumn{3}{c}{Scenario 7} && \multicolumn{3}{c}{Scenario 8}\\
        $b_3$& 0.45 & 0.50 & 0.55 && 0.20 & 0.25 & 0.35 && \textbf{0.30} & 0.35 & 0.40 && 0.15 & \textbf{0.30} & 0.45\\
        $b_2$& 0.25 & \textbf{0.30} & 0.40 && 0.10 & 0.15 & \textbf{0.30} && 0.05 & 0.20 & 0.25 && 0.10 & 0.25 & 0.40\\
        $b_1$& 0.15 & 0.20 & 0.35 && 0.01 & 0.03 & 0.05 && 0.01 & 0.10 & 0.15 && 0.05 & 0.20 & 0.35\\\hline

        & \multicolumn{3}{c}{Scenario 9} && \multicolumn{3}{c}{Scenario 10} && \multicolumn{3}{c}{Scenario 11} && \multicolumn{3}{c}{Scenario 12}\\
        $b_3$& 0.10 & 0.25 & \textbf{0.30} && 0.40 & 0.55 & 0.70 && \textbf{0.30} & 0.50 & 0.60 && 0.45 & 0.50 & 0.60\\
        $b_2$& 0.05 & 0.15 & 0.20 && \textbf{0.30} & 0.45 & 0.60 && 0.20 & 0.40 & 0.55 && \textbf{0.30} & 0.40 & 0.55\\
        $b_1$& 0.01 & 0.03 & 0.07 && 0.20 & \textbf{0.30} & 0.50 && 0.10 & \textbf{0.30} & 0.45 && 0.10 & 0.20 & \textbf{0.30}\\\hline

        & \multicolumn{3}{c}{Scenario 13} && \multicolumn{3}{c}{Scenario 14} && \multicolumn{3}{c}{Scenario 15} && \multicolumn{3}{c}{Scenario 16}\\
        $b_3$& 0.45 & 0.50 & 0.60 && \textbf{0.30} & 0.40 & 0.60 && 0.25 & \textbf{0.30} & 0.50 && \textbf{0.30} & 0.50 & 0.60\\
        $b_2$& 0.10 & \textbf{0.30} & 0.40 && 0.15 & 0.20 & 0.50 && 0.10 & 0.20 & 0.40 && 0.10 & \textbf{0.30} & 0.40\\
        $b_1$& 0.05 & 0.20 & \textbf{0.30} && 0.05 & 0.10 & \textbf{0.30} && 0.05 & 0.15 & \textbf{0.30} && 0.05 & 0.20 & 0.25\\\hline

        & \multicolumn{3}{c}{Scenario 17} && \multicolumn{3}{c}{Scenario 18} && \multicolumn{3}{c}{Scenario 19}\\
        $b_3$& \textbf{0.30} & 0.40 & 0.60 && 0.15 & \textbf{0.30} & 0.50 && \textbf{0.30} & 0.40 & 0.60\\
        $b_2$& 0.15 & 0.20 & \textbf{0.30} && 0.10 & 0.20 & \textbf{0.30} && 0.15 & \textbf{0.30} & 0.50\\
        $b_1$& 0.05 & 0.10 & 0.25 && 0.01 & 0.05 & 0.25 && 0.10 & 0.20 & \textbf{0.30}\\\hline
    \end{tabular}
    \caption{True toxicity scenario, MTCs are highlighted in \textbf{bold}. \label{Tab: scenarios}}
\end{table}

The non-parametric partial ordering benchmark (PO-benchmark) gives an upper bound of the PCS under any combination-escalation designs with binary toxicity endpoint~\citep{MozgunovPavel2020Abfd}. We compare the PCS of the POCRM under the above two ordering specification choices with the PO-benchmark, and the ratios are given below each PCS in Table~\ref{Tab: motivation 3x3}. The POCRM under the 2 choices of orderings generally give PCS comparable to the benchmark, the ratios are between 0.9 and 1.1. However, with Wages 6 orderings, the POCRM has particularly low PCS under scenario 5, only achieving 58\% of the benchmark PCS. On the other hand, with the consistent 6 orderings, the PCS is similar to that given by the PO-benchmark under scenario 5. As it will be shown below, this observation is not specific to these 9 scenarios, but holds in general: the POCRM with Wages 6 orderings always tend to give low PCS when the true MTC is in the middle of the combination grid. 
\begin{table}[!hbt]
    \centering
    \begin{tabular}{l ccccccccc r}
        \hline & \multicolumn{9}{c}{Scenario}\\ \cline{2-10}
        & 1 & 2 & 3 & 4 & 5 & 6 & 7 & 8 & 9 & Mean\\ \hline
        Benchmark & 64.1 & 26.7 & 24.7 & 28.9 & 26.1 & 26.9 & 21.6 & 23.6 & 64.9 & 31.2\\ \hline
        Wages 6 orderings & 59.7 & 34.8 & 20.4 & 32.2 & 15.1 & 26.7 & 19.3 & 27.4 & 71.8 & 30.2\\
        (Ratio) & 0.93 & 1.30 & 0.83 & 1.11 & \textbf{0.58} & 0.99 & 0.89 & 1.16 & 1.11\\ \hline
        Consistent 6 orderings & 60.1 & 32.7 & 22.1 & 33.9 & 26.0 & 28.5 & 16.2 & 22.2 & 72.4 & 31.2\\
        (Ratio) & 0.94 & 1.22 & 0.89 & 1.17 & 1.00 & 1.06 & 0.75 & 0.94 & 1.12\\\hline
    \end{tabular}
    \caption{PCS (\%) under the POCRM with Wages 6 orderings and consistent 6 orderings, the nonparametric benchmark, and the ratio between the two based on $10^4$ simulations. \label{Tab: motivation 3x3}}
\end{table}

\begin{figure}[!hbt]
    \centering
    \includegraphics[width=0.8\linewidth]{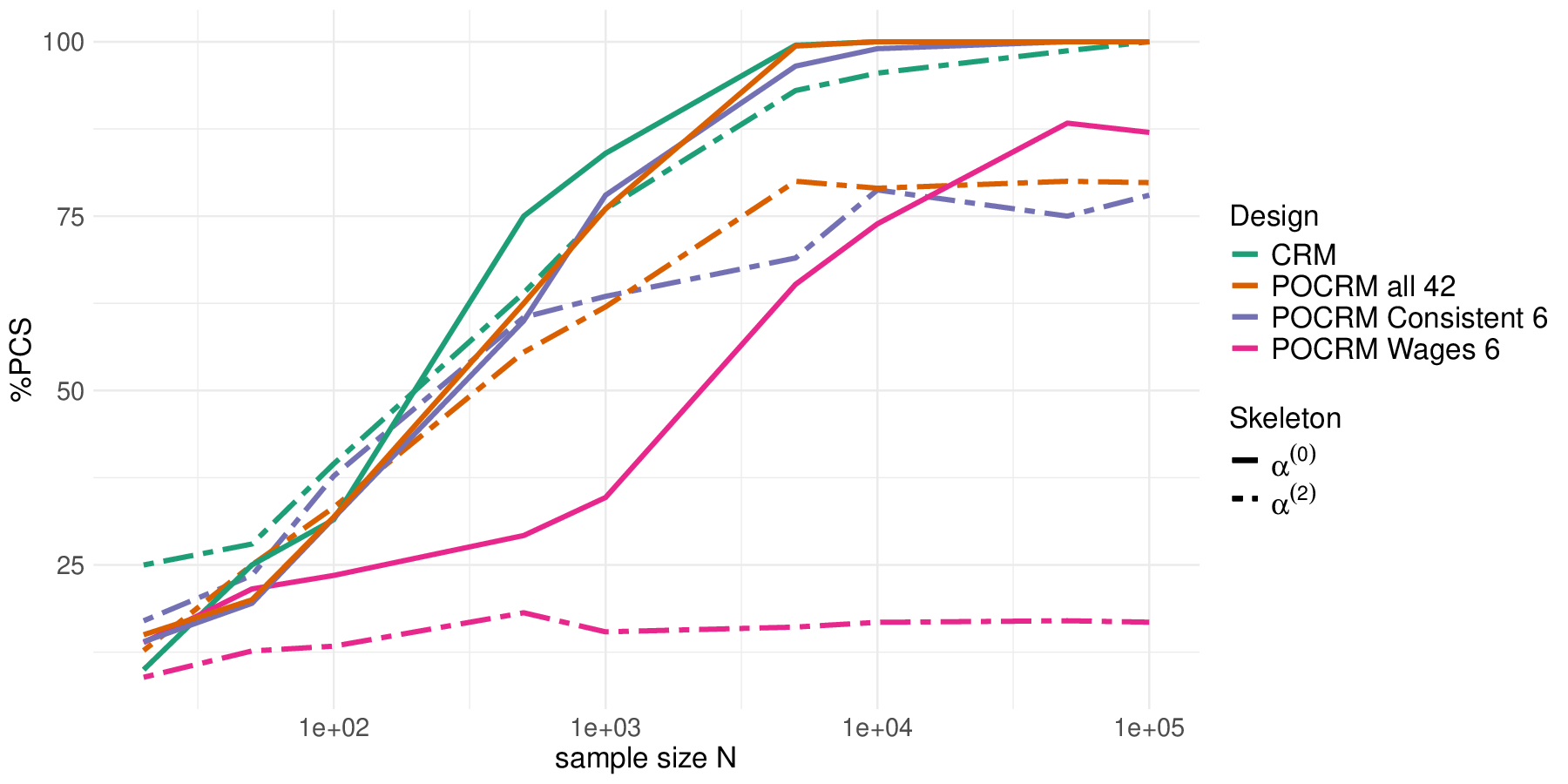}
    \caption{\%PCS vs. $N$ under the CRM with the correct ordering (green), the POCRM with all 42 orderings (orange), and the POCRM with Wages 6 orderings (purple) under skeletons $\bm{\alpha}^{(0)}$ (dashed) and the consistent $\bm{\alpha}^{(2)}$ (solid). Estimations based on $10^4$ simulations.\label{Fig: motivation}}
\end{figure}

Furthermore, the asymptotic behaviour under scenario 5 is investigated in Figure~\ref{Fig: motivation}. Under skeleton $\bm{\alpha}^{(0)}$ (shown by dashed lines), the PCS under the CRM with the correct ordering converges to 100\%, whereas, the PCS under the POCRM does not converge to 100\% under either all 42 orderings, Wages or consistent 6 orderings. This shows that the consistency of the CRM under the correct ordering does not guarantee the consistency of the POCRM, even with the correct ordering included. Then, under a consistent skeleton (elaborated in Section~\ref{sec: calibration}) $\bm{\alpha}^{(2)}$=(0.25, 0.28, 0.34, 0.36, 0.40, 0.44, 0.47, 0.53, 0.55), the PCS under the CRM, the POCRM with 42 orderings, and the POCRM with consistent 6 orderings converge to 100\%, whereas, the PCS under the POCRM with Wages 6 orderings fails to converge to 100\%. This shows that even if consistency can be achieved with all orderings, it is not guaranteed with inappropriate choice of orderings. 

\section{Background}
\label{Sec: CRM POCRM intro}

\subsection{The continual reassessment method (CRM)}
\label{Subsec: CRM}
Following the notations used in~\cite{Shen1996Consistency}, consider a single-agent trial of one treatment with ordered dose levels $\tilde{d}_1<\cdots<\tilde{d}_k$. The aim is to locate the MTD with a \textit{target toxicity level} (TTL) $\theta_0$. Assume the relationship between toxicity and dose levels is monotonically increasing. Let $\psi(\tilde{d}_i,a)$ be the toxicity probability of $\tilde{d}_i$, under the power model $\psi(\tilde{d}_i,a)=\{\alpha[i]\}^a$, where $a\in\mathcal{A}$ is the model parameter and the \textit{toxicity skeleton} $\alpha[i]$ are fixed prior estimates of the toxicity probability at doses $\tilde{d}_i$, $i=1,\ldots,k$.

Assume the toxicity endpoint is binary, with 1 denoting a dose-limiting toxicity (DLT) and 0 a non-DLT. Patients are enrolled in cohorts of $m$. The CRM has two stages. In stage 1, patients are entered into a sequence of pre-specified doses. Stage 1 ends when the first heterogeneity exists. For stage 2, let $Y_1,\ldots,Y_N$ be the number of toxicities in the $j=1,\ldots,N$th cohort. Let $x_j$ be the dose level given to cohort $j$, $x_j\in\{d_1,...,d_k\}$. Let $\Omega_j$ denote the information set after enrolling cohort $j$, i.e. $\Omega_j=\{x_1,y_1,...,x_{j}, y_{j}\}$. Then, the likelihood is binomial, $\mathcal{L}(a;\Omega_j)=\prod_{l=1}^{j}\mathrm{Bin}(\,y_l;m,\psi(x_l,a)\,)$. This is maximised in the interior of $\mathcal{A}$ as soon as there exists heterogeneity in the data, i.e. when there are both toxic and non-toxic events. Then, the model parameter is estimated as the maximum likelihood estimator (MLE)
$\hat{a}_j=\arg\max_{a\in\mathcal{A}}\mathcal{L}(a;\Omega_j)$. The toxicity probability of each dose is estimated by plugging in the MLE $\hat{p}_i=\psi(\tilde{d}_i,\hat{a}_j)$, and the $(j+1)$th cohort will then be allocated to the dose level closest to $\theta_0$ based on the current best guess, i.e. $x_{j+1}=x_{i^\ast}$, where $i^\ast=\arg\min_{i=1,...,k}|\hat{p}_i-\theta_0|$. The final recommended MTD will be $x_{N+1}$.

\subsection{The consistency of CRM}
\label{Subsec: CRM consistency}
\cite{Shen1996Consistency} provided a sufficient condition under which the CRM is consistent. More explicitly, let $x_0$ denote the true MTC, consistency refers to
$x_{N+1}\to x_0$ as $N\to\infty$.
Let $R(d)=\mathrm{pr}(Y=1|x=d)$ be the true toxicity probability at dose $d$, and $R_i=R(\tilde{d}_i)$, $i=1,\ldots,k$. We consider working models $\psi$ satisfy the following assumptions. (M1) $\forall a$, $\psi(\cdot,a)$ is strictly increasing. $a\mapsto\psi(x,a)$ is continuous and strictly monotone in the same direction $\forall x$. (M2) $\forall t\in(0,1)$, $\forall x$, the function $s(t,x,a)=t\frac{\psi'}{\psi}(x,a)+(1-t)\frac{-\psi'}{1-\psi}(x,a)$ is continuous and strictly monotone in $a$. (M3) $a$ belongs to a finite interval $\mathcal{A}=[a_1,a_{m+1}]$. The three assumptions on the true toxicity probability $R$ are (D1) $R(x_0)=\theta_0$. (D2) $0<R_1<\cdots<R_k<1$. (D3) $S=\{a:|\psi(x_0,a)-\theta_0|<|\psi(x_i,a)-\theta_0|,\ \forall x_i\neq x_0\}$. Let $\psi(x_i,a_i)=R_i$, $a_i\in S$, $\forall i$. \cite{Shen1996Consistency} show that the CRM is consistent given (M1)-(M3) and (D1)-(D3). Nevertheless, condition (D3) can be restrictive and~\cite{Cheung2002RelaxedConsistency} postulate a relaxed condition, which will be referred as the \textit{CRM consistency condition}.

\begin{lemma}[CRM consistency condition]
Suppose $d_l$ is the MTC. Given Assumptions (M1)-(M3), (D1)-(D2), CRM is consistent if 
\begin{equation*}
a_l\in H_l,\qquad
a_i\in\bigcup_{j=i+1}^k H_j,\quad i=1,\ldots,l-1,\qquad
a_i\in\bigcup_{j=1}^{i-1}H_j,\quad i=l+1,\ldots,k,
\end{equation*}
where $\mathcal{A}=H_1\cup H_2\cup\cdots\cup H_{k+1}$, $H_1=[b_1, b_2)$, $H_i=(b_i, b_{i+1})$, $i=2,\ldots,m-1$, $H_k=(b_k, b_{k+1}]$, and $\psi(d_{i-1},b_i)+\psi(d_{i},b_i)=2\theta_0$, $a_i=\log R_i/\log\alpha[i]$, $i=2,...,k$. 
\label{Lemma: CRM consistency}
\end{lemma}

\subsection{The partial ordering CRM (POCRM)}
\label{Subsec: POCRM}
Consider the combination of two agents $A, B$, each with levels $a_1<\cdots<a_r$, $b_1<\cdots<b_c$, respectively. The goal is to find the \textit{maximum tolerated combination} (MTC). Let $\tilde{d}_{i,j}=(a_i,b_j)$, $i=1,\ldots,r$, $j=1,\ldots,c$, $k=rc$, which can be arranged into a matrix as Table~\ref{Tab: 2x2 grid}. 
\begin{table}[!hbt]
\centering
\begin{tabular}{c c c}\hline
    & \multicolumn{2}{c}{Drug A}\\ \cline{2-3}
    Drug & $\tilde{d}_{1,2}=(a_1,b_2)$ & $\tilde{d}_{2,2}=(a_2,b_2)$\\
    B & $\tilde{d}_{1,1}=(a_1,b_1)$ & $\tilde{d}_{2,1}=(a_2,b_1)$\\ \hline
\end{tabular}
\caption{Example of $2\times 2$ dose-combination grid.\label{Tab: 2x2 grid}}
\end{table}

The $2\times2$ example in Table~\ref{Tab: 2x2 grid} gives partial orderings $\tilde{d}_{1,1}\to \tilde{d}_{2,1}\to \tilde{d}_{2,2}$ and $\tilde{d}_{1,1}\to \tilde{d}_{1,2}\to \tilde{d}_{2,2}$, resulting in 2 complete orderings: Ordering 1: $\tilde{d}_{1,1}\to \tilde{d}_{2,1}\to \tilde{d}_{1,2}\to \tilde{d}_{2,2}$; Ordering 2: $\tilde{d}_{1,1}\to \tilde{d}_{1,2}\to \tilde{d}_{2,1}\to \tilde{d}_{2,2}$. Due to this uncertainty in orderings, the CRM in Section~\ref{Subsec: CRM} cannot be applied directly. \cite{Wages2011TwoStageCRM}~suggests using \textit{Bayesian model selection} to select the complete ordering most compatible with data. Let $\psi_m(\tilde{d}_{i,j},a)=\{\alpha_{m}(i,j)\}^a$ be the CRM under ordering $m$, $m=1,\ldots,M$, where $\bm{\alpha}_{m}$ is an re-ordered version of the monotonically increasing toxicity skeleton $\bm{\alpha}$ with respect to ordering $m$. The POCRM design has an additional ordering selection step than the CRM. Stage 1 remains the same. In stage 2, after $n$ patients are evaluated, the MLE of $a$ under ordering $m$ is $\hat{a}_n^{(m)}$, where the likelihood $\mathcal{L}_m(a|\Omega_n)\propto\prod_{l=1}^n\psi_m\{x_l,a\}^{y_l}\left(1-\psi_m\{x_l,a\}\right)^{1-y_l}$.
Let $p(m)$ be the prior probability of ordering $m$, the posterior probability is defined as
\begin{equation}
p(m|\Omega_n)=\frac{\exp\left\{\mathcal{L}_m(\hat{a}_n^{(m)}|\Omega_n)\right\}p(m)}{\sum_{m'=1}^M\exp\left\{\mathcal{L}_{m'}(\hat{a}_n^{(m')}|\Omega_n)\right\}p(m')}.
\label{Eqn: post prob model}
\end{equation}
ordering $m^\ast$, which maximises~\eqref{Eqn: post prob model}, is chosen. The CRM is applied under $m^\ast$ to select the combination $x_{n+1}$ assigned to the next cohort.

\section{The consistency of POCRM}
\label{Sec: Consistency derivation}
The reason for the undermined performance of POCRM, as illustrated in Section~\ref{Subsec: motivation}, is that the design is not asymptotically consistent. Hence, in this section, a set of sufficient conditions for the consistency of the POCRM has been provided. We start from the simplest case of $2\times2$-combination in Section~\ref{Sec: POCRM consistency 2x2}, the key is to provide conditions that ensures the selection of the correct ordering. The $2\times3$-combination case in Section~\ref{Subsec: consistency 2x3} is used to introduce the idea of ``correct ordering group”, which is central to the guidance on the consistency conditions for orderings. Section~\ref{Subsec: consistency general} generalises the conditions to the $r\times c$ case, and finally, Section~\ref{Subsec: check condition 2x2} provides guidance on checking the conditions. Proofs of all theorems, lemmas and proposition are provided in Supplementary Materials.

\subsection{$2\times2$ dose-combination setting}
\label{Sec: POCRM consistency 2x2}
Consider $2\times2$-combinations as in Table~\ref{Tab: 2x2 grid}. Assume ordering 1 is correct and the CRM under ordering 1 is consistent. Both orderings are included into the POCRM.

As the sample size grows, as long as neither ordering has prior probability 0, the comparison of posteriors goes to the likelihoods. The one with the largest maximised log-likelihood
\[\ell_m\left(\hat{a}_n^{(m)}|\Omega_n\right)=\left.\sum_{i=1}^r\sum_{j=1}^c\tilde{n}_{i,j}\left[\frac{\tilde{y}_{i,j}}{\tilde{n}_{i,j}}\log\{\alpha_{m}(i,j)\}^a+\left(1-\frac{\tilde{y}_{i,j}}{\tilde{n}_{i,j}}\right)\log(1-\{\alpha_{m}(i,j)\}^a)\right]\right|_{a=\hat{a}_n^{(m)}},\]
is selected, where $\tilde{y}_{i,j}=\sum_{l=1}^n y_l \mathbbm{1}\{x_l=\tilde{d}_{i,j}\}$, and $\tilde{n}_{i,j}=\sum_{l=1}^n\mathbbm{1}\{x_l=\tilde{d}_{i,j}\}$.
Re-scaling the log-likelihoods by the sample size to make it stable as $n\to\infty$:
\begin{equation*}
\frac{\ell_m\left(\hat{a}_n^{(m)}|\Omega_n\right)}{n}=\left.\sum_{i=1}^r\sum_{j=1}^c\eta_{i,j}\left[\frac{\tilde{y}_{i,j}}{\tilde{n}_{i,j}}\log\{\alpha_{m}(i,j)\}^a+\left(1-\frac{\tilde{y}_{i,j}}{\tilde{n}_{i,j}}\right)\log(1-\{\alpha_{m}(i,j)\}^a)\right]\right|_{a=\hat{a}_n^{(m)}},
\label{Eqn: rescaled ll}
\end{equation*}
where $\eta_{i,j}=\tilde{n}_{i,j}/n$ is the proportion of patients at $\tilde{d}_{i,j}$. Following~\cite{Shen1996Consistency}, the proportion of toxicity at each $d_{i,j}$ converges to the true toxicity probability $R_{i,j}$, and thus, the MLE $\hat{a}_n^{(m)}\to\hat{a}^{(m)}$, the (re-scaled) log-likelihood converges to
\begin{equation*}
\tilde{\ell}_m:=\left.\sum_{i,j}\eta_{i,j}\left[R_{i,j}\log\{\alpha_{m}(i,j)\}^a+(1-R_{i,j})\log(1-\{\alpha_{m}(i,j)\}^a)\right]\right|_{a=\hat{a}^{(m)}}
\end{equation*}

With 2 orderings, the focus is the difference between the (re-scaled) log-likelihoods $\Delta\ell=\tilde{\ell}_1-\tilde{\ell}_2$. Ordering~1 is selected if $\Delta\ell\geq0$. Then, since the CRM is consistent under ordering~1, the PCS will converge to 1 to give consistency. The consistency condition for $2\times2$ combinations is summarised in Theorem~\ref{Theorem: 2x2 general} below.
\begin{theorem}[POCRM consistency in the $2\times2$ case]
For $2\times2$ combinations, assume all complete orderings are included and the true ordering is $\tilde{d}_{1,1}\to\tilde{d}_{2,1}\to\tilde{d}_{1,2}\to\tilde{d}_{2,2}$, the POCRM is consistent given the following sufficient conditions:
\begin{itemize}
\item The prior probabilities of both orderings are non-zero.
\item If $\tilde{d}_{1,1}$ or $\tilde{d}_{2,2}$ is the MTC, the CRM under both orderings is consistent.
\item If $\tilde{d}_{2,1}$ or $\tilde{d}_{1,2}$ is the MTC, the CRM is consistent under the correct ordering, and
\begin{align*}
    f(\{\alpha[1]\}^{\hat{a}^{(1)}}, R_{1,1})\geq f(\{\alpha[1]\}^{\hat{a}^{(2)}}, R_{1,1}) \text{ when }\eta_{2,1}\text{ small}, & \text{if }\tilde{d}_{1,2}\text{ MTC};\\
    f(\{\alpha[4]\}^{\hat{a}^{(1)}}, R_{2,2})\geq f(\{\alpha[4]\}^{\hat{a}^{(2)}}, R_{2,2}) \text{ when }\eta_{1,2}\text{ small}, & \text{if }\tilde{d}_{1,2}\text{ MTC}.
\end{align*}
\end{itemize}
Where, $\alpha[l]$, $l=1,\ldots,4$, is a monotonically increasing skeleton, $(\eta_{i,j})_{i,j=1}^2$ and $(R_{i,j})_{i,j=1}^2$ are the proportion of patients and true toxicity probabilities at $\tilde{d}_{i,j}$. $\hat{a}^{(1)}$ and $\hat{a}^{(2)}$ are converged values of MLEs under ordering 1 and 2, and $f(\hat{R}, R)=R\log\hat{R}+(1-R)\log(1-\hat{R})$.
\label{Theorem: 2x2 general}
\end{theorem}
The notion ``small" in Theorem~\ref{Theorem: 2x2 general} is defined as follows.
\begin{definition}
For $2\times2$ combinations with $\tilde{d}_{2,1}$ being the MTC and ordering 1 being correct, the proportion of patients assigned to $\tilde{d}_{2,1}$, $\eta_{2,1}$, is \textbf{small} when the converged values of MLEs $\hat{a}^{(1)}>\hat{a}^{(2)}$. Similarly, if $\tilde{d}_{1,2}$ is the MTC, $\eta_{1,2}$ is small when $\hat{a}^{(1)}<\hat{a}^{(2)}$.
\label{Def: eta small}
\end{definition}

\subsection{$2\times 3$ dose-combination setting}
\label{Subsec: consistency 2x3}

Consider now $2\times3$ combinations with three levels of drug A and two of drug B, as in Table~\ref{Tab: 2x3 grid}. We introduce the ideas of the \textit{relabelling} of combinations, and the \textit{correct ordering groups}, which allow the assumption of including all orderings to be relaxed. 
\begin{table}[!hbt]
\centering
\begin{tabular}{l c c c}\hline
    & \multicolumn{3}{c}{Drug A}\\ \cline{2-4}
    Drug & $\tilde{d}_{1,2}$=($a_1,b_2$) & $\tilde{d}_{2,2}$=($a_2,b_2$) & $\tilde{d}_{3,2}$=($a_3,b_2$)\\
    B & $\tilde{d}_{1,1}$=($a_1,b_1$) & $\tilde{d}_{2,1}$=($a_2,b_1$) & $\tilde{d}_{3,1}$=($a_3,b_1$)\\ \hline
\end{tabular}
\caption{$2\times3$ dose-combinations.\label{Tab: 2x3 grid}}
\end{table}

\begin{definition}[Relabelling of combinations, single MTC]
    Define the relabelling operator $\mathfrak{L}$, which maps the combination $\tilde{d}_{i,j}$ to its label $d_l$ given the scenario of true toxicity probabilities $\mathbf{R}$, $i=1,\ldots,r$; $j=1,\ldots,c$; $l=1,\ldots,k=rc$, such that there is only one single MTC. Upon relabelling, the toxicity probabilities increase in the order $d_1\to d_2\to\ldots\to d_k$.
    \label{Def: relabelling}
\end{definition}

\begin{definition}[correct ordering group]
    Under a given scenario $\mathbf{R}$ with a single MTC $\tilde{d}_{i^\ast,j^\ast}$, let $d_\nu=\mathfrak{L}(\tilde{d}_{i^\ast,j^\ast};\mathbf{R})$ be the label of the MTC upon relabelling per Definition~\ref{Def: relabelling}. Let $\mathcal{O}[l]$ be the $l$th combination under ordering $\mathcal{O}$, and $\mathcal{O}[1:l]=\{\mathcal{O}[1], \mathcal{O}[2],\ldots,\mathcal{O}[l]\}$. Then, ordering $\mathcal{O}$ belongs to the correct ordering group of $\mathbf{R}$ if and only if
    \begin{enumerate}
        \item $\mathcal{O}[\nu]=d_\nu=\tilde{d}_{i^\ast,j^\ast}$, i.e. the MTC is ordered correctly;
        \item $\mathcal{O}[1:(\nu-1)]=\{d_1,\ldots,d_{\nu-1}\}$, i.e. all combinations less toxic than the MTC are ordered before the MTC.
    \end{enumerate}
    \label{Def: group}
\end{definition}

For example, under scenario $\mathbf{R}=\begin{pmatrix}
    0.3 & 0.4 & 0.6\\
    0.1 & 0.2 & 0.5
\end{pmatrix}$, the combinations are relabelled as $\mathfrak{L}(\tilde{d};\mathbf{R})=\begin{pmatrix}
    d_3 & d_4 & d_6\\
    d_1 & d_2 & d_5
\end{pmatrix}$, so that the least toxic combination $\tilde{d}_{1,1}$ gets label $d_1$, the second least toxic combination $\tilde{d}_{2,1}$ gets $d_2$, etc. For $2\times3$ cases, there are 5 possible orderings:
\begin{equation}
    \begin{split}
        &\mathcal{O}_1: \tilde{d}_{1,1}\to \tilde{d}_{2,1}\to \tilde{d}_{3,1}\to \tilde{d}_{1,2}\to \tilde{d}_{2,2}\to \tilde{d}_{3,2}; \,\, \mathcal{O}_2: \tilde{d}_{1,1}\to \tilde{d}_{1,2}\to \tilde{d}_{2,1}\to \tilde{d}_{2,2}\to \tilde{d}_{3,1}\to \tilde{d}_{3,2};\\
        &\mathcal{O}_3: \tilde{d}_{1,1}\to \tilde{d}_{2,1}\to \tilde{d}_{1,2}\to \tilde{d}_{3,1}\to \tilde{d}_{2,2}\to \tilde{d}_{3,2}; \,\, \mathcal{O}_4: \tilde{d}_{1,1}\to \tilde{d}_{1,2}\to \tilde{d}_{2,1}\to \tilde{d}_{3,1}\to \tilde{d}_{2,2}\to \tilde{d}_{3,2};\\
        &\mathcal{O}_5: \tilde{d}_{1,1}\to \tilde{d}_{2,1}\to \tilde{d}_{1,2}\to \tilde{d}_{2,2}\to \tilde{d}_{3,1}\to \tilde{d}_{3,2}.
    \end{split}\label{Eqn: 2x3 orderings}
\end{equation}
Suppose the MTC is $\tilde{d}_{1,2}$ with TTL $\theta=0.30$, then the label of the MTC is $\mathfrak{L}(\tilde{d}_{1,2};\mathbf{R})=d_3$, ($\nu=3$ in Definition~\ref{Def: group}). Orderings 3 and 5 satisfy the first condition with $\mathcal{O}_3[3]=\mathcal{O}_5[3]=\tilde{d}_{1,2}$, whereas the other 3 orderings order the MTC incorrectly. For condition 2, both ordering 3 and 5 have $\mathcal{O}_3[1:2]=\mathcal{O}_5[1:2]=\{\tilde{d}_{1,1},\tilde{d}_{2,1}\}=\{d_1,d_2\}$, the two combinations less toxic than the MTC are ordered before the MTC. Hence, orderings 3 and 5 belong to the correct ordering group, whereas orderings 1, 2, 4 are incorrect under scenario $\mathbf{R}$. 

As the number of complete ordering grows, instead of operating on orderings, we now operate on \textit{ordering groups}. Selecting any ordering in the correct group leads to recommending the true MTC. The assumption of including the correct ordering can therefore be relaxed. The POCRM consistency can be achieved as a direct generalisation of the $2\times2$ case. The following set of notations will be used in Theorem~\ref{Theorem: 2x3 general} and~\ref{Theorem: rxc consistency}.
\begin{definition}[Notations]
    Given the scenario $\mathbf{R}$, for any combination $\tilde{d}_{i,j}$, let $\sigma_m(i,j)$ be the order of $\tilde{d}_{i,j}$ under $\mathcal{O}_m$, i.e. if $s=\sigma_m(i,j)$, $\mathcal{O}_m[s]=\tilde{d}_{i,j}$. Then, denote the true toxicity probability at $s$ as $R_m[s]$, $R_m[s]=R_{i,j}$. Under another ordering $\mathcal{O}_{m'}$, write $\alpha_{m,m'}[s]=\alpha_{m'}(i,j)$. Intuitively, given the $s$th toxic combination under $\mathcal{O}_m$, $\alpha_{m,m'}[s]$ gives its skeleton under $\mathcal{O}_{m'}$. $\alpha[l]$ is the $l$th element of the monotonically increasing skeleton $\bm{\alpha}$.
    \label{Def: notations}
\end{definition}
Under $\mathbf{R}$ above Equation~\eqref{Eqn: 2x3 orderings} for instance, $\sigma_2(2,1)=3$, $\tilde{d}_{2,1}$ is ordered third under $\mathcal{O}_2$. The true toxicity of $\tilde{d}_{2,1}$ is $R_{2,1}=0.2$, which is also denoted as $R_2[3]$, the toxicity of ``the combination ordered third under $\mathcal{O}_2$". Suppose the monotonically increasing skeleton $\bm{\alpha}=(0.1, 0.2, 0.3, 0.4, 0.5, 0.6)$. Then $\alpha[1]=0.1$, $\alpha[3]=0.3$, for instance. Under ordering 2, the re-ordered skeleton is $\bm{\alpha}_2=\begin{pmatrix}
    0.2 & 0.4 & 0.6\\
    0.1 & 0.3 & 0.5
\end{pmatrix}$, and thus $\alpha_2(2,1)=0.3$. Under ordering 1, $\alpha_{2,1}[3]$ asks for the skeleton of ``the third toxic combination under $\mathcal{O}_2$" under $\mathcal{O}_1$, i.e. the skeleton of $\tilde{d}_{2,1}$ under $\mathcal{O}_1$. $\bm{\alpha}_1=\begin{pmatrix}
    0.4 & 0.5 & 0.6\\
    0.1 & 0.2 & 0.3
\end{pmatrix}$, the skeleton of $\tilde{d}_{2,1}$ is 0.2, $\alpha_{2,1}[3]=0.2$.

\begin{theorem}[POCRM consistency in the $2\times3$ case]
Under the $2\times3$ toxicity scenario $\mathbf{R}$ with the MTC $\tilde{d}_{i^\ast,j^\ast}$ labelled as $d_\nu$, the POCRM is consistent given sufficient conditions
\begin{itemize}
    \item Include $\geq1$ orderings, all consistent under CRM, from the correct ordering group $\mathcal{C}$.
    \item If $\nu=1$ or 6, $\mathcal{C}=\{\mathcal{O}_1,\mathcal{O}_2,\mathcal{O}_3,\mathcal{O}_4,\mathcal{O}_5\}$.
    \item If $\nu=2$ or 5, $f^t(\alpha[1],R_{1,1})\geq f^m(\alpha[1],R_{1,1})$ or $f^t(\alpha[6],R_{3,2})\geq f^m(\alpha[6],R_{3,2})$, respectively, $\forall t\in\mathcal{C}, m\notin\mathcal{C}$. 
    \item If $\nu=3$, $\forall t\in\mathcal{C}$, $\mathcal{C}=\{\mathcal{O}_2,\mathcal{O}_4\}$ when $d_3=\tilde{d}_{2,1}$, $f^t(\alpha_{1,t}[5],R_1[5])\geq f^1(\alpha[5],R_1[5])$, $f^t(\alpha_{m,t}[4],R_m[4])\geq f^m(\alpha[4],R_m[4])$, $m=3,5$. $\mathcal{C}=\{\mathcal{O}_3,\mathcal{O}_5\}$ when $d_3=\tilde{d}_{1,2}$, $f^t(\alpha_{1,t}[2],R_{2,1})\geq f^1(\alpha[2],R_{2,1})$, $f^t(\alpha_{m,t}[4],R_m[4])\geq f^m(\alpha[4],R_m[4])$, $m=2,4$. $\mathcal{C}=\{\mathcal{O}_1\}$ when $d_3=\tilde{d}_{3,1}$, $f^t(\alpha_{2,t}[3],R_2[3])\geq f^2(\alpha[3],R_2[3])$, $f^t(\alpha[1],R_{1,1})\geq f^m(\alpha[1], R_{1,1})$, $m=2,4$, $f^t(\alpha[2],R_{2,1})\geq f^m(\alpha[2], R_{2,1})$, $m=3,5$.
    \item If $\nu=4$, $\forall t\in\mathcal{C}$, $\mathcal{C}=\{\mathcal{O}_3,\mathcal{O}_4\}$ when $d_4=\tilde{d}_{3,1}$, $f^t(\alpha[5],R_{2,2})\geq f^1(\alpha[5],R_{2,2})$, $f^t(\alpha_{m,t}[3],R_m[3])\geq f^m(\alpha[3], R_m[3])$, $m=2,5$. $\mathcal{C}=\{\mathcal{O}_2,\mathcal{O}_5\}$ when $d_4=\tilde{d}_{2,2}$, $f^t(\alpha_{1,t}[2],R_{2,1})\geq f^1(\alpha[2],R_{2,1})$, $f^{t}(\alpha_{m,t}[3],R_m[3])\geq f^m(\alpha[3],R_m[3])$, $m=2,4$. $\mathcal{C}=\{\mathcal{O}_1\}$ when $d_4=\tilde{d}_{1,2}$, $f^t(\alpha_{2,t}[4],R_2[4])\geq f^2(\alpha[4],R_2[4])$, $f^t(\alpha[6], R_{3,2})\geq f^m(\alpha[6],R_{3,2})$, $m=2,5$, $f^t(\alpha_{m,t}[5],R_m[5])\geq f^m(\alpha[5], R_m[5])$, $m=3,4$.
\end{itemize}
where the five orderings are as defined in Equation~\eqref{Eqn: 2x3 orderings}, $f^m(\alpha,R)=f(\alpha^{\hat{a}^{(m)}},R)=R\log(\alpha^{\hat{a}^{(m)}})+(1-R)\log(1-\alpha^{\hat{a}^{(m)}})$, all other notations defined in Definition~\ref{Def: notations}.
\label{Theorem: 2x3 general}
\end{theorem}

\subsection{General $r\times c$ case}
\label{Subsec: consistency general}
This section considers the general $r\times c$ case where drug A has $c$ levels and drug B has $r$ levels. Given the toxicity scenario $\mathbf{R}$, let $\tilde{d}_{i^\ast,j^\ast}$ be the true MTC and $d_\nu=\mathfrak{L}(\tilde{d}_{i^\ast,j^\ast};\mathbf{R})$. Algorithm~\ref{Alg: CRM consistency} constructs a skeleton making all orderings in the correct group consistent under the CRM. Under ordering $t$ in the correct group, if $\sigma_t(i,j)<l$ for $d_l=\mathfrak{L}(d_{i,j};\mathbf{R})$, $l<\nu$, i.e. $\tilde{d}_{i,j}$ is less toxic than the MTC and ordered before its correct location under $\mathcal{O}_t$, a lower bound is imposed on $\alpha[l]$. Conversely, if $\sigma_t(i,j)>l$ for $d_l=\mathfrak{L}(\tilde{d}_{i,j};\mathbf{R})$, $l>\nu$, an upper bound is imposed on $\alpha[l]$. The only potential obstacle is that the upper bound for $\alpha[l]$ might be smaller than $\alpha[l-1]$, or vice versa. Lemma~\ref{Lemma: CRM consistency algo} shows these will not happen.
\begin{lemma}
Algorithm \ref{Alg: CRM consistency} gives a skeleton strictly monotonically increasing.
\label{Lemma: CRM consistency algo}
\end{lemma}

\begin{algorithm}[!hbt]
\caption{CRM consistency of the correct group}\label{Alg: CRM consistency}
\begin{algorithmic}[1]
\Require The true MTC $d_\nu$; The correct ordering group; The initial skeleton $\bm{\alpha}$.
\For{Ordering $t$ in the correct group}
	\State $a_{t}(i,j)=\log R_{i,j}/\log\alpha_t(i,j)$, $i=1,\ldots,r$, $j=1,\ldots,c$.
    \State Let $a_t[l]=a_t(i,j)$ for $\mathfrak{L}(\tilde{d}_{i,j}$, $l=1,\ldots,k$.
	\For{$l=(\nu-1)$ to 1}
		\If{$\mathfrak{L}(\tilde{d}_{i,j};\mathbf{R})=d_l$ and $\sigma_t(i,j)<l$}
			\State Increases $\alpha[l]$ to the smallest value such that $$a_t[l]>b_{l+1},\qquad\text{where}\qquad \alpha[l]^{b_{l+1}}+\alpha[l+1]^{b_{l+1}}=2\theta_0.$$
		\EndIf
	\EndFor
	\For{$l=(\nu+1)$ to $k$}
		\If{$\mathfrak{L}(\tilde{d}_{i,j};\mathbf{R})=d_l$ and $\sigma_t(i,j)>l$}
			\State Decreases $\alpha[l]$ to the largest value such that $$a_t[l]<b_{l},\qquad\text{where}\qquad \alpha[l-1]^{b_{l}}+\alpha[l]^{b_{l}}=2\theta_0.$$
		\EndIf
	\EndFor
\EndFor
\end{algorithmic}
\end{algorithm}

\begin{definition}
    Given scenario $\mathbf{R}$ with MTC $\tilde{d}_{i^\ast,j^\ast}$ labelled as $d_\nu$, for ordering $m$ not in the correct group, the set $\mathcal{T}_1(m;\mathbf{R})=\{l=1,\ldots,k: d_l=\mathfrak{L}(\tilde{d}_{i,j};\mathbf{R}), l>\nu, \sigma_m(i,j)<\nu\}$ contains all combinations more toxic than the MTC that are ordered before the MTC under $\mathcal{O}_m$. Let $\mathcal{W}_1(m;\mathbf{R})=\{l=1,\ldots,k:(l+1)\in\mathcal{T}_1(m;\mathbf{R}), l\notin\mathcal{T}_1(m;\mathbf{R})\}$ be the set of combinations ordered immediately before the combinations in $\mathcal{T}_1(m;\mathbf{R})$ but are not in $\mathcal{T}_1(m;\mathbf{R})$ themselves. Similarly, let $\mathcal{T}_2(m;\mathbf{R})=\{l=1,\ldots,k: d_l=\mathfrak{L}(\tilde{d}_{i,j};\mathbf{R}), l<\nu, \sigma_m(i,j)>\nu\}$ be all combinations less toxic than the MTC that are ordered after the MTC, and $\mathcal{W}_2(m;\mathbf{R})=\{l=1,\ldots,k:(l-1)\in\mathcal{T}_2(m;\mathbf{R}), l\notin\mathcal{T}_2(m;\mathbf{R})\}$ be the combinations ordered immediately after them. Let $\mathcal{W}(m;\mathbf{R})=\mathcal{W}_1(m;\mathbf{R})\cup \mathcal{W}_2(m;\mathbf{R})$.
    \label{Def: relevant combo}
\end{definition}

\begin{theorem} [POCRM consistency condition, $r\times c$ case]
For $r\times c$ dose-combinations, if at least one ordering, ordering $t$, in the correct group $\mathcal{C}$ under scenario $\mathbf{R}$ is included, the POCRM is consistent given the following sufficient conditions.
\begin{enumerate}
    \item The CRM under $\mathcal{O}_t$ is consistent, $\forall t\in\mathcal{C}$.
    \item Define $f^m(\alpha,R)=f(\alpha^{\hat{a}^{(m)}},R)=R\log(\alpha^{\hat{a}^{(m)}})+(1-R)\log(1-\alpha^{\hat{a}^{(m)}})$,
        \begin{equation}
            f^t\left(\alpha_{m,t}[l],R_m[l]\right)\geq f^m\left(\alpha[l],R_m[l]\right),
            \quad l\in\mathcal{W}(m;\mathbf{R}), t\in\mathcal{C}, m\notin \mathcal{C},
            \label{Eqn: general consistency condition}
        \end{equation}
        where $\mathcal{W}$ is defined in Definition~\ref{Def: relevant combo} and all other notations defined in Definition~\ref{Def: notations}.
\end{enumerate}
\label{Theorem: rxc consistency}
\end{theorem}

\subsection{Check POCRM consistency conditions}
\label{Subsec: check condition 2x2}
The CRM consistency ensures that for orderings in the correct group $\mathcal{C}$, the MLE converges to $\hat{a}^{(t)}=\log (R_{i^\ast,j^\ast})/\log(\alpha[\nu])$, $\forall\mathcal{O}_t\in\mathcal{C}$ when $d_\nu=\tilde{d}_{i^\ast,j^\ast}$. However, for $\mathcal{O}_m\notin\mathcal{C}$, the converged value $\hat{a}^{(m)}$ depends on the proportion of patient allocated to each combination, $\eta_{i,j}$'s. In practice, $\eta_{i,j}$'s are obtained by simulations, which is computationally costly. Hence, one solution is to consider all possible $\eta_{i,j}$'s, and show that condition~\eqref{Eqn: general consistency condition} holds for all $\eta_{i,j}$'s. 

For all $\eta_{i,j}$ such that $\sum_{i,j}\eta_{i,j}=1$, the converged value of MLEs solves
\begin{equation}
\left.\sum_{i=1}^r\sum_{j=1}^c\eta_{i,j}(1-R_{i,j})\log\alpha_{m}(i,j)\left(\frac{R_{i,j}}{1-R_{i,j}}-\frac{\{\alpha_{m}(i,j)\}^a}{1-\{\alpha_{m}(i,j)\}^a}\right)\right|_{a=\hat{a}^{(m)}}=0,\quad m\notin\mathcal{C}.
\label{Eqn: converged value of MLEs}
\end{equation}
Then, $\hat{a}^{(m)}$ are used in Theorem~\ref{Theorem: rxc consistency} to check the POCRM consistency. 

\subsection{Multiple MTCs}
It has been assumed so far that the true scenario has a single MTC. In practice, it is of interest to look for a maximum tolerated contour on which all combinations have toxicity equal to the TTL. The design is \textit{consistent} if the total probability of selecting any MTCs converges to 1 as $N\to\infty$. The relabelling technique is generalised in Definition~\ref{Def: relabelling with multiple MTCs} below. Each MTC obtains the label $d_\nu$ in exactly one version of the labels. The definition of correct ordering group remains the same as Definition~\ref{Def: group}, and the POCRM is consistency as long as Theorem~\ref{Theorem: rxc consistency} is satisfied under any one version of the relabelling.
\begin{definition}[Relabelling with multiple MTCs]
    Given the toxicity scenario $\mathbf{R}$, let $\tilde{d}_{i_1^\ast,j_1^\ast}$, $\tilde{d}_{i_2^\ast,j_2^\ast}$, $\ldots$, $\tilde{d}_{i_P^\ast,j_P^\ast}$ be the MTCs for some $P\in\mathbb{N}$. Let $(\nu-1)$ be the number of combinations with toxicity probabilities less than the MTC. Then, the combinations are relablled $P$ time, resulting in $P$ versions of the label $\mathbf{d}^{(p)}=\mathfrak{L}^{(p)}(\tilde{\mathbf{d}};\mathbf{R})$, $p=1,\ldots,P$ such that
    \begin{enumerate}
        \item $\mathfrak{L}^{(p)}(\tilde{d}_{i_p^\ast,j_p^\ast};\mathbf{R})=d_\nu$, $p=1,\ldots,P$;
        \item The toxicity probability is non-decreasing following the ordering $d_1\to d_2\to\ldots\to d_k$.
    \end{enumerate}
    \label{Def: relabelling with multiple MTCs}
\end{definition}

\subsection{Consistency under multiple scenarios}
The consistency condition in Theorem~\ref{Theorem: rxc consistency} assumed a single known toxicity scenario $\mathbf{R}$, whereas in practice, the true scenario is unknown and the design is often evaluated under various scenarios that are likely to approximate the truth. This section gives a necessary condition for the POCRM to be consistent across various scenarios. Since the consistency under scenarios with multiple MTCs can be implied from scenarios with a single MTC, the focus would be the latter case. Explicitly, consider $k$ scenario $\mathbf{R}^{(1)}, \ldots, \mathbf{R}^{(k)}$ such that the MTC under $\mathbf{R}^{(l)}$ is labelled $d_l$, $l=1,\ldots,k$. One necessary part of the POCRM consistency is that the CRM has to be consistent under the correct ordering, which gives the following. 
\begin{proposition}[Consistency under multiple scenarios]
    Under scenarios with single MTC $\mathbf{R}^{(l)}$, $l=1,\ldots,k$ such that the MTC under $\mathbf{R}^{(l)}$ is labelled $d_l$ per Definition~\ref{Def: relabelling}, let $\mathcal{O}_{l}$ be the correct ordering under $\mathbf{R}^{(l)}$. A necessary condition for the POCRM to be consistent under all $k$ scenarios simultaneously is, for $i=2,\ldots,k-1$,
    \begin{equation}
        \begin{split}
        \max\{R^{(2)}_{2}[1],..., R^{(k)}_{k}[1]\}<\alpha[1]^{b_2}<R^{(1)}_{1}[1];&\quad R^{(k)}_{k}[k]<\alpha[k]^{b_k}<\min\{R^{(1)}_{1}[k], ..., R^{(k-1)}_{k-1}[k]\};\\
        \max\{R^{(i+1)}_{i+1}[i],..., R^{(k)}_{k}[i]\}<\alpha[i]^{b_{i+1}}<&R^{(i)}_{i}[i]<\alpha[i]^{b_i}<\min\{R^{(1)}_{1}[i],..., R^{(i-1)}_{i-1}[i]\},
        \end{split}
    \label{Eqn: multiple scen}
    \end{equation}
    \label{Prop: multiple scen}
    where $b_1,\ldots,b_{k+1}$ are as defined in Lemma~\ref{Lemma: CRM consistency}, all other notations defined in Definition~\ref{Def: notations}.
\end{proposition}

\section{Ordering specification}
\label{Subsec: number of models}
The number of orderings increases combinatorially with the levels of agents. For $2\times3$ combinations, there are 5 orderings, which increases to 42 for $3\times 3$, 462 for $3\times4$, and 24024 for $4\times4$ cases, which makes it impractical to include all orderings. The orderings included should cover the true ordering, be computationally efficient, and reflect clinical considerations. Below, we focus solely on statistical considerations and refer readers to~\cite{Mozgunov2022PracticalImplementation} for guidance on ordering specification in collaboration with clinicians. 

\cite{Wages2013Orders}~suggest 6 orderings (\textit{Wages 6 orderings}), which gives good operational characteristics. Nevertheless, the example in Section~\ref{Subsec: motivation} shows that this choice leads to very low accuracy under certain scenarios. Hence, we propose a novel method to specify orderings motivated by the consistency condition that at least one ordering from the correct group should be included. \textit{Scenario-specific} choice of orderings assumes the set of scenarios are known, while this set can contain as many scenarios as needed to cover all interesting cases. \textit{Scenario-agnostic} assumes the true scenario is unknown and the specified orderings should be consistent under any toxicity scenario. The full list of possible orderings is assumed known and we select subsets from the list.

\subsection{Scenario-specific ordering specification}
\begin{figure}[!hbt]
    \centering
    \includegraphics[width=0.65\linewidth]{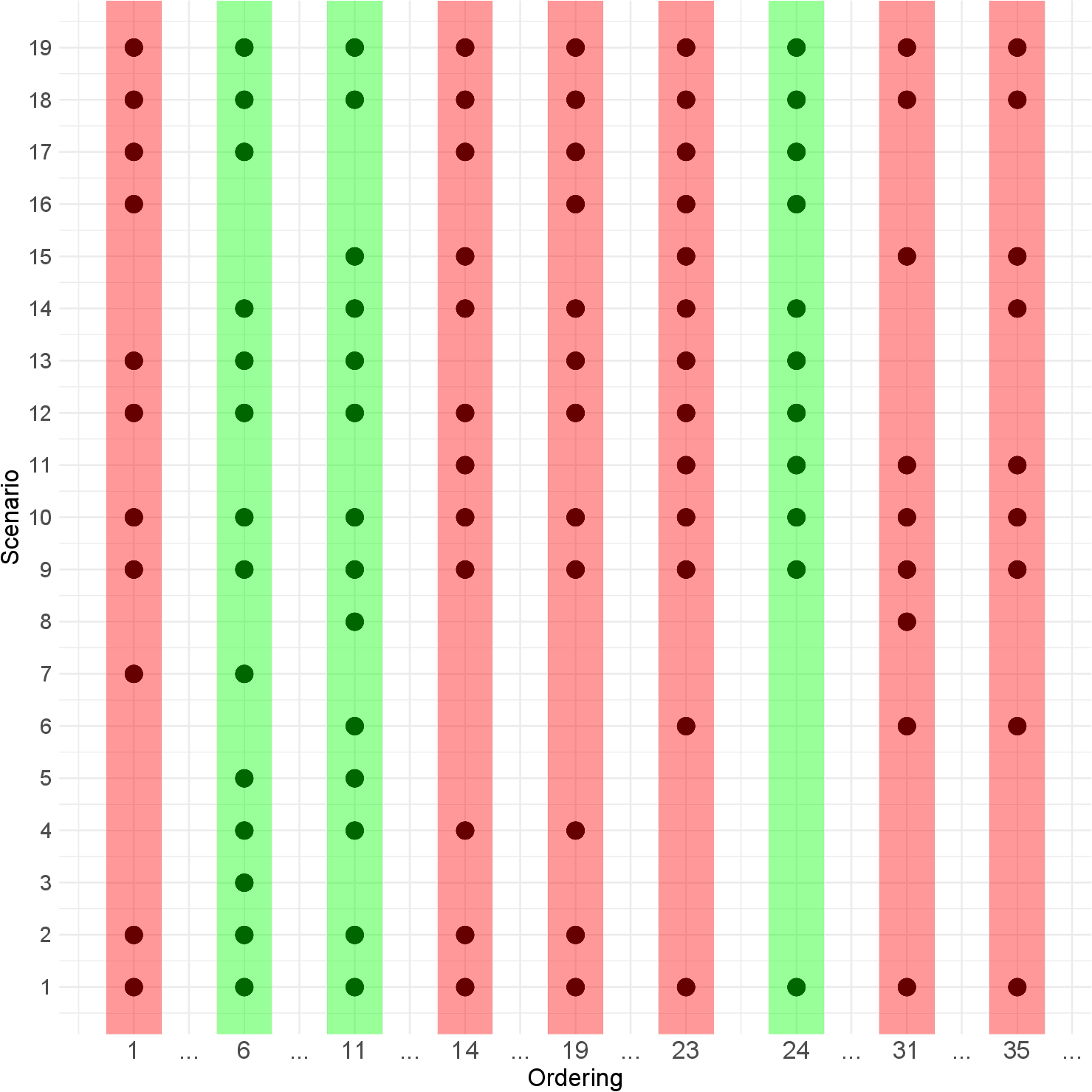}
    \caption{Dot at $(m,r)$ means $\mathcal{O}_m$ belongs to the correct ordering group under scenario $r$, $m=1,\ldots,42; r=1,\ldots,19$. Green and red are consistent and inconsistent choices.\label{Fig: correct group}}
\end{figure}

Figure~\ref{Fig: correct group} gives an example of scenario-specific specification under the 19 scenarios in Table~\ref{Tab: scenarios}. For $3\times3$ combinations, there are 42 possible orderings, listed in Supplementary materials. The 19 scenarios are shown on the y-axis and selected orderings are shown on the x-axis. A dot at coordinate $(m,r)$ means ordering $m$ belongs to the correct group under scenario $r$, $m=1,\ldots,42$, $r=1,\ldots,19$. For example, the first column tells that ordering 1 belongs to the correct groups under scenarios 1, 2, 7, 9, 10, 12, 13, 16-19. The last column tells that ordering 35 belongs to the correct group under scenarios 1, 6, 9-11, 14, 15, 18, 19. 

 Consistency requires that among the included orderings (columns), the correct group under each scenario should be covered at least once, which translates to at least one dot on each row among the selected columns. The smallest number of columns to achieve this goal is 3, for example, orderings 6, 11, 24 highlighted in green. On the other hand, the Wages 6 orderings corresponds to orderings 1, 14, 19, 23, 31, 35, highlighted in red. Among these 6 columns, there is no dot on scenario 5, i.e. no ordering in the correct group under scenario 5 has been included. Hence, the POCRM with Wages 6 orderings cannot be consistent under scenario 5, which explains the undermined performance described in Section~2 (Table~\ref{Tab: motivation 3x3}). 

\subsection{Scenario-agnostic ordering specification}
Nevertheless, the above scenario-specific ordering specification method relies on the assumed combination-toxicity scenarios. In practice, the true toxicity scenario is unknown, and it would be beneficial to have an ordering specification method that ensures consistency regardless of the true toxicities. The key is to realise that although the number of possible toxicity scenarios is infinite, it is \textit{order-scenario}, defined below, that characterises the correct ordering group, and the number of order-scenarios is finite.
\begin{definition}[Order-scenario]
    Let $\mathbf{R}^{(1)}, \mathbf{R}^{(2)}$ be two toxicity scenarios with a same single MTC $\tilde{d}_{i^\ast,j^\ast}$. Let $d_{\nu_1}=\mathfrak{L}(\tilde{d}_{i^\ast,j^\ast};\mathbf{R}^{(1)})$, $d_{\nu_2}=\mathfrak{L}(\tilde{d}_{i^\ast,j^\ast};\mathbf{R}^{(2)})$ be the labels of the MTC. Define the set of combinations with toxicity probabilities below the MTC $B_c=\{\tilde{d}_{i,j}:\mathfrak{L}(\tilde{d}_{i,j};\mathbf{R}^{(c)})=d_l, l<\nu_c\}$, $c=1,2$, referred to as the ``below MTC set". Then,  $\mathbf{R}^{(1)}$ and $\mathbf{R}^{(2)}$ belong to the same order-scenario if and only if $\nu_1=\nu_2$ and $B_1=B_2$.
    \label{Def: order scenario}
\end{definition}
\begin{proposition}[Correct ordering groups are characterised by order-scenarios]
    Given two toxicity scenarios $\mathbf{R}^{(1)}$ and $\mathbf{R}^{(2)}$ belonging to the same order-scenario. If ordering $\mathcal{O}$ belongs to the correct group under $\mathbf{R}^{(1)}$, then it also belongs to the correct group under $\mathbf{R}^{(2)}$.
    \label{Prop: group depend on order scen}
\end{proposition}

Hence, one way to specify orderings regardless of the true toxicity scenario is to list out all order-scenarios and ensure at least one ordering in the correct group has been included under each possible order-scenario. Note that the order-scenario is defined by 3 elements, the MTC $\tilde{d}_{i^\ast,j^\ast}$, the label of the MTC $\nu$, and the below MTC set $B$. Moreover, when specific toxicity scenarios are not assumed, the consistency of scenarios with multiple MTCs are implied by the single MTC scenarios, and thus we focus only on the single MTC case below.

\begin{figure}[!hbt]
    \centering
    \includegraphics[width=0.9\linewidth]{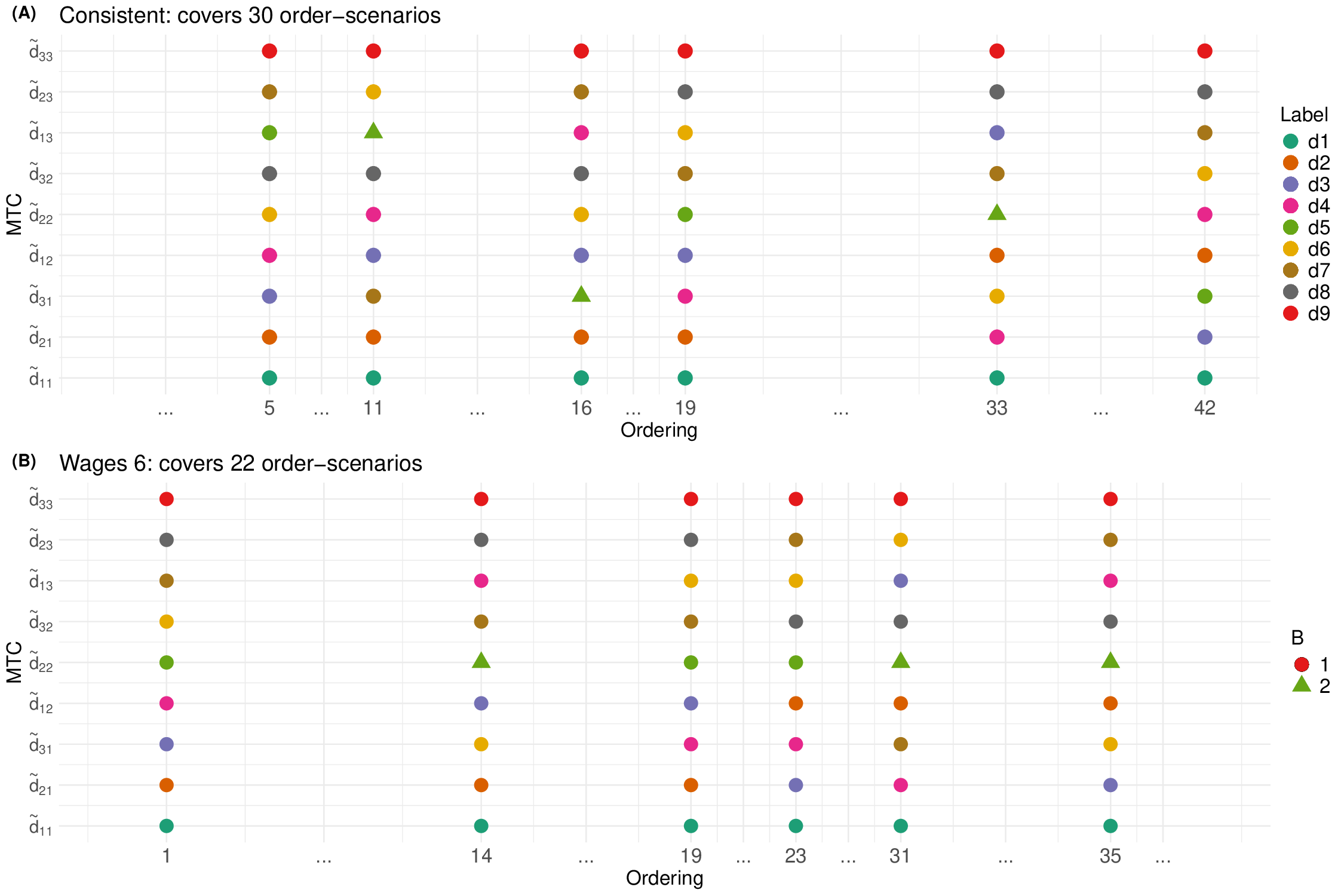}
    \caption{All possible correct group under all orderings and all positions of the MTC. Positions of MTCs by colours and groups by shapes of the dots.\label{Fig: all correct groups}}
\end{figure}

For $3\times3$ combinations, there are 30 possible order-scenarios, which is graphically shown in Figure~\ref{Fig: all correct groups}. The 3 elements that defines an order-scenario are shown by the y-axis (the MTC), colours (the label of the MTC, $\nu$), and point types (the below MTC set $B$). For example, the dark green dots on the bottom row corresponds to an order-scenario where the MTC is $\tilde{d}_{1,1}$, labelled as $d_1$, in which case, there is only 1 possible below MTC set, $\emptyset$. The light green dots on the third row from the bottom in (Panel A) corresponds to the order-scenario where the MTC is $\tilde{d}_{3,1}$, labelled as $d_5$, in which case there are 2 possible below MTC sets, shown by the triangles and circles. The x-axis of each dot gives the ordering in the correct group under this order-scenario. Hence, the included orderings (columns) should together cover all 30 order-scenarios. The consistent 6 orderings, ordering 5, 11, 16, 19, 33, 42, shown in (Panel A) satisfies this criterion, whereas the Wages 6 orderings shown in (Panel B) does not satisfy this criterion. In particular, all dots in the middle row (when MTC is $\tilde{d}_{2,2}$) are light green in (Panel B), which reveals that the Wages 6 orderings completely missed the possibility that the MTC is labelled as $d_5$. In fact, scenario 5 in Table~\ref{Tab: motivation 3x3} belongs to the order-scenario represented by the pink dot in the middle row, i.e. $\tilde{d}_{2,2}$ is labelled $d_4$, and hence Wages 6 orderings does not lead to a consistent design. Six is the smallest number of orderings that satisfies the consistency condition, but the choice of six is not unique. One possible way to find the optimal six is discussed in Section~\ref{Subsec: efficiency}.

\section{Calibration of model parameters}
\label{sec: calibration}
\subsection{Amend toxicity skeletons}
\label{Subsec: check consistency conditions}
Figure~\ref{Fig: motivation} shows that under skeleton $\bm{\alpha}^{(0)}$=(0.10, 0.20, 0.30, 0.40, 0.45, 0.50, 0.54, 0.59, 0.64), the CRM under the correct ordering is consistent while the POCRM is not, even under all 42 orderings, because the consistency condition in Equation~\eqref{Eqn: general consistency condition} is not satisfied. This section illustrates the way to check the consistency condition and suggests an amendment method if not satisfied, using scenario 5 in Table~\ref{Tab: scenarios} as an example.

The first step is to relabel the combinations and find the correct group. Given scenario 5, 
$\mathfrak{L}(\tilde{\mathbf{d}};\mathbf{R}^{(5)})=\begin{pmatrix}
    d_7 & d_8 & d_9\\
    d_3 & \mathbf{d_4} & d_6\\
    d_1 & d_2 & d_5
\end{pmatrix}$.
The correct ordering group contains ordering 6-11, 37-42. The condition of including at least one ordering from the correct group is satisfied trivially with all 42 orderings, but not satisfied by the Wages 6 orderings. Furthermore, the CRM has to be consistent under any of the 12 orderings in the correct group, which is true only for orderings 6-11, but not 37-42 under $\bm{\alpha}^{(0)}$. Specifically, the condition $a_2=\log R^{(5)}_{3,1}/\log\alpha^{(0)}[2]\in (b_3,\infty)$ is violated, where $b_3$ solves $\alpha^{(0)}[2]^{b_3}+\alpha^{(0)}[3]^{b_3}=2\theta_0$, all notations as defined in Definition~\ref{Def: notations}. Increasing $\alpha^{(0)}[2]$ from 0.20 to 0.21 fixes this condition.

Next, orderings 1-3, 12-14, 19-24, 31-36 have the MTC labelled $d_5$, and the set $\mathcal{W}$ per Definition~\ref{Def: relevant combo} contains either $\tilde{d}_{2,1}=d_2$ or $\tilde{d}_{1,2}=d_3$. The true probabilities are $R^{(5)}_{2,1}=0.2$, $R^{(5)}_{1,2}=0.25$. Hence, the condition in Equation~\eqref{Eqn: general consistency condition} translates to
\begin{equation}
    \begin{split}
        f^6(\alpha[2],0.2)\geq f^m(\alpha[2],0.2),\qquad & m=1-3, 34-36;\\
        f^6(\alpha[3],0.25)\geq f^m(\alpha[3],0.25),\qquad & m=12-14, 19-24, 31-33.
    \end{split}
    \label{Eqn: condition scen5}
\end{equation}
Note that the function $f$ implicitly depends on the proportion of patients assigned to each combinations, $\eta_{i,j}$, which is not available without simulations. One possible solution is to require Equation~\eqref{Eqn: condition scen5} to be satisfied under all possible values of $\eta_{i,j}$ s.t. $\sum_{i,j}\eta_{i,j}=1$. However, this is computationally expensive if $k$ is large. We suggest to take a subset of the combinations that contains the MTC, the set $\mathcal{W}$, and their neighbours under both orderings considered. Take $m=1$ for example, under the correct ordering 6, $\mathcal{W}=\emptyset$, only the MTC $d_4$ and its two neighbours $d_2,d_3$ are considered. Under the incorrect ordering 1, $\mathcal{W}(2;\mathbf{R}^{(5)})=\{d_2\}$, with neighbours $d_1$ and $d_5$ ($\mathcal{O}_1$ is $d_1\to d_2\to d_5\to d_3\to d_4\to\cdots$). Hence, the subset of combinations to be considered are $d_1,d_2,d_3,d_4,d_5$. We randomly draw $5\times10^4$ values of $\eta_{i,j}$ such that $\sum_{l=1}^5\eta_l=1$ where $\eta_l$ is the proportion of patients assigned to $d_l$, and check~\eqref{Eqn: condition scen5}. Repeat this procedure for all orderings 1-3, 12-14, 19-24, 31-36, the consistency condition requires increasing $\alpha[2]$ from 0.21 to 0.25, and $\alpha[3]$ from 0.30 to 0.32.

Similarly, for orderings 4, 5, 15-18, 25-30, the MTC is labelled $d_6$ and the set $\mathcal{W}$ can be $\{d_2\}$, $\{d_3\}$, or $\{d_2,d_3\}$. The POCRM consistency condition in Equation~\eqref{Eqn: general consistency condition} translates to 
\begin{align*}
    &f^6(\alpha[2], 0.2)\geq f^m(\alpha[2],0.2), \quad f^6(\alpha[3],0.25)\geq f^m(\alpha[4],0.25),\quad  m=4,5;\\
    &f^6(\alpha[3],0.25)\geq f^m(\alpha[3],0.25),\,\,  m=15-18;\quad 
    f^6(\alpha[2],0.2)\geq f^m(\alpha[3],0.2),\,\,  m=25-28;\\
    &f^6(\alpha[2],0.2)\geq f^m(\alpha[4],0.2), \quad f^6(\alpha[3],0.25)\geq f^m(\alpha[2],0.25),\quad  m=29,30.
\end{align*}
This requires adjusting $\alpha[2]$ to 0.27, $\alpha[4]$ to 0.37. Thus, $\bm{\alpha}^{(1)}$=(0.10, 0.27, 0.32, 0.37, 0.45, 0.50, 0.54, 0.59, 0.64) makes the POCRM with all 42 orderings consistent under scenario~5.

\subsection{Consistency under multiple scenarios}
\label{Subsec: multiple scen}
The skeleton $\bm{\alpha}^{(1)}$ makes the POCRM consistent under scenario 5 but not all scenarios, this section goes on to check the other scenarios. The consistency under scenarios 10-19 with multiple MTCs would be implied by the single MTC scenario, and thus we focus on scenarios 1-9 with a single MTC. Upon relabelling, the condition in Equation~\eqref{Eqn: multiple scen} is satisfied if we define scenario 3, 4, 5, 6, 8 as $\mathbf{R}^{(5)}, \mathbf{R}^{(3)}, \mathbf{R}^{(4)}, \mathbf{R}^{(8)}, \mathbf{R}^{(6)}$, respectively. Under $\alpha^{(1)}$, the values of $b_i$'s are $\{b_i\}_{i=2}^k$=(0.70, 0.99, 1.13, 1.35, 1.62, 1.84, 2.11, 2.48). Scenario~1 violates $(0.40=)\alpha^{(1)}[2]^{b_2}<R_{2,1}^{(1)}=0.35$; Scenario~2 violates $(0.25=)R_{1,1}^{(2)}<\alpha^{(1)}[1]^{b_2}(=0.20)$; and Scenario~4 violates $(0.20=)R_{1,1}^{(3)}<\alpha^{(1)}[1]^{b_1}(=0.19)$. Algorithm~\ref{Alg: CRM consistency} suggests increasing $\alpha[1]$ to 0.13, which makes the CRM consistent under the correct ordering groups in all 9 scenarios.

Furthermore, Equation~\eqref{Eqn: general consistency condition} are checked. Under scenario 2, incorrect orderings have the MTC labelled $d_3$ and the set $\mathcal{W}=\{d_1\}$. Consistency requires $f^t(\alpha[1],0.25)\geq f^m(\alpha[1],0.25$ for ordering $t$ in the correct group $\mathcal{C}$ and $m\notin\mathcal{C}$, which increases $\alpha[1]$ to 0.18. Under scenario 4, incorrect orderings have the MTC labelled $d_2$ and $\mathcal{W}=\{d_4\}$ or $\{d_5\}$. The consistency condition would require $f^t(\alpha[4],0.35\geq f^m(\alpha[l],0.35$ and $f^t(\alpha[5],0.4)\geq f^m(\alpha[l],0.4)$, for $l=4,5$, and $f^t(\alpha[6],0.45)\geq f^m(\alpha[4],0.45)$. This requires $\alpha[4]\leq0.36, \alpha[5]\leq0.40$. The other scenarios are checked similarly, and the final skeleton is $\bm{\alpha}^{(2)}=$(0.25, 0.28, 0.34, 0.36, 0.40, 0.44, 0.47, 0.53, 0.55), which is the consistent skeleton used in Figure~\ref{Fig: motivation}.

\subsection{Efficiency}
\label{Subsec: efficiency}
As mentioned in Section~\ref{Subsec: number of models}, the choice of consistent orderings might not be unique. One way to choose a set of orderings to use in the study is to assess its PCS at small sample sizes. The term efficiency is defined as the mean PCS across all scenarios of consideration at $N=60$. This section chooses orderings such that the efficiency can be maximised.

Intuitively, an ordering is good if it belongs to the correct group under a large number of scenarios. This translates to maximising the number of dots in the selected columns in Figure~\ref{Fig: correct group}. Motivated by this, the metric \verb|n.consis| is defined to be the total number of dots in the selected columns standardised by the number of orderings selected, $S$. \texttt{n.consis}=$\left(\sum_{c=1}^C\sum_{m=1}^S\mathbbm{1}\{\mathcal{O}_m\in\text{correct group of }\mathbf{R}^{(c)}\}\right)/S$, where $\mathrm{R}^{(1)},\ldots,\mathbf{R}^{(C)}$ are the scenarios considered. For example, the $S=3$ consistent orderings highlighted in green in Figure~\ref{Fig: correct group} has 39 dots over the $C=19$ scenarios, which gives \verb|n.consis|=13. The Wages 6 orderings highlighted in red has 66 dots, and thus \verb|n.consis|=11. Table~\ref{Tab: efficiency} summarises the \verb|n.consis| and the mean PCS according to the number of orderings, $S$, and whether the choice is consistent. The left part gives the mean \verb|n.consis| and PCS based on 50 different random choices of orderings in the corresponding category, and the right gives, within this category, the choice with the best and worst PCS.
If the number of possible choice is less than 50, all choices are used. The s.e. gives the standard error of the estimated PCS in 19 scenarios. The skeleton $\bm{\alpha}^{(0)}$ is used and all estimates are based on $10^4$ simulations. 
\begin{table}[!hbt]
    \centering
    \begin{tabular}{l c c c c ccc }
        \hline S & Consis? & \verb|n.consis| & PCS (s.e.) && Ordering & \texttt{n.consis} & PCS\\ \hline
                                                                                          &&&&& 6, 11, 24 & 13.0 & 46.72 (17.29)\\
         & \multirow{-2}{*}{Yes} & \multirow{-2}{*}{11.98} & \multirow{-2}{*}{45.43 (17.10)} && 1, 8, 31 & 11.3 & 44.26 (16.88)\\ \cline{2-8}
         &&&&& 6, 7, 11 & 14.7 & 48.48 (17.00)\\
         \multirow{-4}{*}{3} & \multirow{-2}{*}{No} & \multirow{-2}{*}{11.10} & \multirow{-2}{*}{41.76 (20.21)} && 24, 25, 28 & 9.7 & 38.60 (23.81)\\ \hline
         &&&&& 6, 7, 11, 13 & 14.0 & 46.59 (16.43)\\
         & \multirow{-2}{*}{Yes} & \multirow{-2}{*}{11.98} & \multirow{-2}{*}{45.68 (17.29)} && 6, 11, 13, 28 & 12.0 & 43.83 (17.92)\\ \cline{2-8}
         &&&&& 6, 7, 11, 12 & 14.0 & 48.31 (16.33)\\
         \multirow{-4}{*}{4} & \multirow{-2}{*}{No} & \multirow{-2}{*}{11.25} & \multirow{-2}{*}{43.85 (19.30)} && 24, 25, 28, 29 & 9.3 & 37.69 (23.76)\\ \hline
         &&&&& 6, 7, 12, 21, 23 & 13.4 & 46.79 (17.37)\\
         & \multirow{-2}{*}{Yes} & \multirow{-2}{*}{11.83} & \multirow{-2}{*}{45.10 (17.82)} && 6, 9, 12, 19, 28 & 11.8 & 44.37 (18.25)\\ \cline{2-8}
         &&&&& \small{7, 8, 19, 35, 41} & 13.0 & 45.65 (18.91)\\
         \multirow{-4}{*}{5} & \multirow{-2}{*}{No} & \multirow{-2}{*}{11.02} & \multirow{-2}{*}{42.04 (19.63)} && \small{24, 25, 28, 29, 33} & 9.2 & 38.59 (22.82)\\ \hline
         &&&&& \small{7, 8, 11, 19, 21, 24} & 13.2 & 47.14 (18.31)\\
         & \multirow{-2}{*}{Yes} & \multirow{-2}{*}{11.48} & \multirow{-2}{*}{44.58 (18.05)} && \small{1, 8, 12, 16, 23, 28} & 11.7 & 44.61 (18.21)\\ \cline{2-8}
         &&&&& \small{7, 9, 11, 19, 27, 41} & 13.0 & 45.94 (18.54)\\
         &&&&& Wages 6 & 11.0 & 43.34 (18.38)\\
         \multirow{-5}{*}{6} & \multirow{-3}{*}{No} & \multirow{-3}{*}{11.20} & \multirow{-3}{*}{43.72 (19.11)} && \small{24, 25, 28, 29, 33, 34} & 9.2 & 38.46 (23.11)\\ \hline
         42 & Yes & 11.29 & 44.84 (17.73)\\\hline
    \end{tabular}
    \caption{\texttt{n.consis} and mean PCS according to the number of orderings, $S$, and consistency.\label{Tab: efficiency}}
\end{table}

The following observations are made. Firstly, consistent orderings give better mean PCS than inconsistent ones in all categories. This confirms that although consistency is an asymptotic property, it plays an important role on the efficiency. Individually, there are cases when the best inconsistent choice gives higher PCS than the best consistent choice, but note that there are much more inconsistent choices than the consistent ones, and hence it is conceivable to have one abnormally good inconsistent choice. Nevertheless, the worst inconsistent choices have PCS much lower than the consistent ones. In particular, the differences between the best and worst consistent choice are around 2\% in all categories, but for inconsistent choices, the differences are around 10\%. Secondly, among consistent choices, the PCS are generally larger for smaller $S$. Although the differences are small, minimising the number of orderings also leads to cheaper computation. Finally, the metric \texttt{n.consis} is almost perfectly correlated with the PCS, all choices with larger PCS are associated with larger \texttt{n.consis}. Hence, the overall recommendation for ordering specification is to choose the one with the highest \texttt{n.consis} and smallest $S$ among consistent choices. Under the 19 scenarios specifically, the recommended choice is ordering 6, 11, 24.

Finally, we compare the PCS using the inconsistent skeleton $\bm{\alpha}^{(0)}$ with the consistent $\bm{\alpha}^{(2)}$, and 4 choices of orderings - all 42 orderings, Wages 6 orderings, scenario-agnostic consistent 6 orderings (ordering 5, 11, 16, 19, 33, 42), and scenario-specific 3 orderings (ordering 6, 11, 24). Table~\ref{Tab: Efficiency PCS} gives the PCS at sample size $N=60$. It shows that consistent orderings does help with the efficiency at small sample sizes, with the scenario-specific 3 orderings giving the best (geometric) mean PCS among the 4 choices under each skeleton. On the other hand, using a consistent skeleton does hinder the efficiency slightly in exchange to the asymptotic convergence to 100\% PCS. 

\begin{table}[!hbt]
    \centering
    \begin{tabular}{l cc c cc c cc c cc}
        \hline & \multicolumn{4}{c}{Inconsistent $\bm{\alpha}^{(0)}$} && \multicolumn{4}{c}{Consistent $\bm{\alpha}^{(2)}$}\\ \cline{2-5}\cline{7-10}
        & All 42 & Wages6 & Agnostic6 & Specific3 && All 42 & Wages6 & Agnostic6 & Specific3\\ \hline
        1 & 69.4 & 59.7 & 60.1 & 62.2 &  & 68.9 & 69.4 & 70.3 & 69.6 \\ 
        2 & 27.0 & 34.8 & 32.7 & 36.8 &  & 28.6 & 28.4 & 27.5 & 29.6 \\ 
        3 & 23.4 & 20.4 & 22.1 & 25.6 &  & 22.9 & 15.8 & 24.5 & 23.3 \\ 
        4 & 31.8 & 32.2 & 33.9 & 34.4 &  & 25.2 & 28.2 & 29.7 & 30.2 \\ 
        5 & 25.2 & 15.1 & 26 & 29.5 &  & 24.7 & 25.0 & 24.2 & 28.2 \\ 
        6 & 25.1 & 26.7 & 28.5 & 27.2 &  & 29.2 & 25.9 & 26.8 & 27.4 \\ 
        7 & 17.1 & 19.3 & 16.2 & 23.2 &  & 17.8 & 17.4 & 16.6 & 21.5 \\ 
        8 & 26.1 & 27.4 & 22.2 & 29.7 &  & 27.7 & 30.2 & 26.7 & 33.4 \\ 
        9 & 71.7 & 71.8 & 72.4 & 70.3 &  & 74.2 & 72.6 & 70.3 & 74.0 \\ 
        10 & 74.0 & 74.2 & 70.8 & 77.0 &  & 62.4 & 62.7 & 62.6 & 67.5 \\ 
        11 & 62.4 & 63.6 & 60.2 & 48.8 &  & 52.5 & 54.6 & 50.6 & 40.0 \\ 
        12 & 62.6 & 64.8 & 64.8 & 58.5 &  & 48.9 & 51.1 & 51.7 & 46.0 \\ 
        13 & 63.6 & 54.8 & 63.2 & 69.4 &  & 52.6 & 55.7 & 56.0 & 63.5 \\ 
        14 & 50.2 & 48.6 & 47.6 & 55.4 &  & 54.5 & 44.7 & 46.0 & 53.2 \\ 
        15 & 49.3 & 51.2 & 48.5 & 47.8 &  & 48.6 & 49.6 & 47.7 & 48.3 \\ 
        16 & 50.4 & 50.8 & 52.2 & 50.4 &  & 50.0 & 52.8 & 46.4 & 50.8 \\ 
        17 & 48.4 & 45.8 & 46.5 & 56.0 &  & 52.5 & 49.5 & 49.8 & 57.9 \\ 
        18 & 64.6 & 64.4 & 63.2 & 64.4 &  & 65.0 & 63.3 & 65.7 & 61.4 \\ 
        19 & 71.7 & 70.8 & 72.7 & 71.4 &  & 69.2 & 69.0 & 70.2 & 69.0 \\ \hline
        Mean & 43.70 & 42.64 & 43.37 & 46.13 && 42.29 & 41.41 & 41.71 & 43.74\\\hline
    \end{tabular}
    \caption{PCS at $N=60$. All estimates based on $10^4$ estimations.\label{Tab: Efficiency PCS}}
    \label{tab:my_label}
\end{table}

\begin{figure}[!htb]
    \centering
    \includegraphics[width=0.75\linewidth]{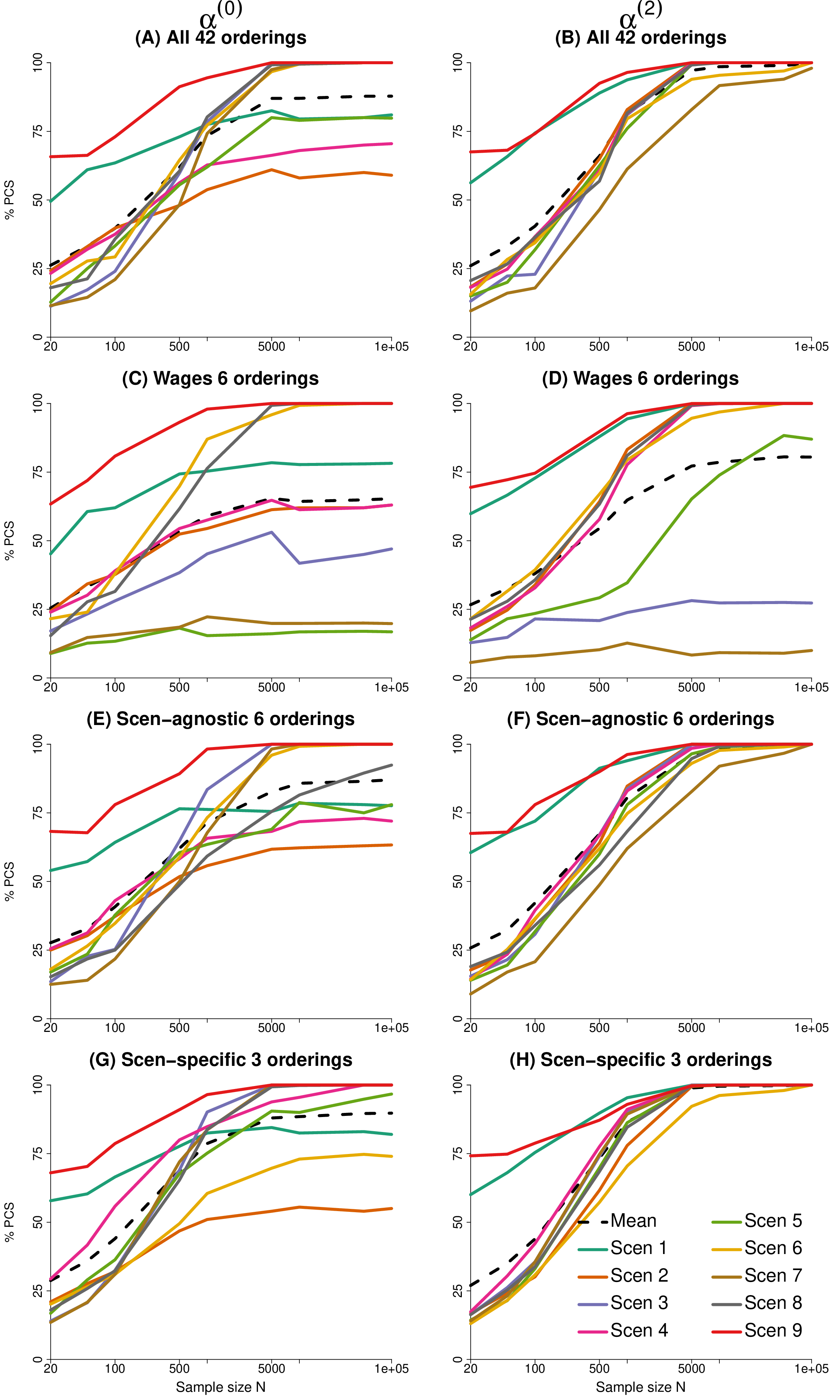}
    \caption{PCS vs.N. Skeletons: $\bm{\alpha}^{(0)}$ (left), $\bm{\alpha}^{(1)}$ (right). Ordering: all 42 (A-B), Wages 6 (C-D), scen-agnostic 6 (E-F), scen-specific 3 (G-H), under scen 1-9 based on $10^4$ simulations. \label{Fig: multiple scen amended}}
\end{figure}

Figure~\ref{Fig: multiple scen amended} looks at the asymptotic behaviour of the POCRM under the above 4 choices of orderings. The PCS is plotted against sample sizes $N$ which ranges from 20 to $10^5$. The inconsistent skeleton $\bm{\mathbf{\alpha}^{(0)}}$ and the consistent skeleton $\bm{\mathbf{\alpha}^{(2)}}$ are used for the left and right column, respectively. Scenarios 1-9 in Table~\ref{Tab: scenarios} with a single MTC are considered, and results for scenarios 10-19 are given in the Supplementary materials, all based on $10^4$ simulations. The first row (Panel A-B) includes all 42 orderings. $\bm{\mathbf{\alpha}^{(0)}}$ is inconsistent under scenario 1, 2, 4, 5 with PCS converges to 65\%-75\%. These are exactly the ones violating Equation~\eqref{Eqn: multiple scen}. Under $\bm{\mathbf{\alpha}^{(2)}}$ (Panel B), the POCRM is consistent in all scenarios. Under Wages 6 orderings (Panel C-D), scenarios 3 and 7, in addition to 1, 2, 4, are inconsistent under $\bm{\alpha}^{(0)}$. This is as expected since Wages 6 orderings does not include any ordering in the correct group. Under $\bm{\alpha}^{(2)}$, scenarios 1, 2, 4 are fixed and 3, 5 and 7 remain inconsistent, since the correct group is still uncovered. For the scenario-agnostic 6 orderings, under $\bm{\alpha}^{(0)}$ (Panel E), the PCS behaves similarly as in (Panel A). Under $\bm{\alpha}^{(2)}$ (Panel F), with only 6 orderings, the POCRM is still consistent under all scenarios. The scenario-specific consistent 3 orderings in (Panel G-H), shows similar trends as in (Panel E-F). This confirms that for the specified 19 scenarios, it suffices to include only 3 orderings to achieve consistency.

\section{Discussion}
\label{Sec: conclusion}
This paper suggests a way to conduct ordering selections for the POCRM in the setting of dual-agent combination-escalation trials. The selection is based on the consistency of the POCRM, which is an asymptotic property that considers whether the design selects the correct MTC with probability one when the sample size goes to infinity. A set of sufficient conditions has been provided to ensure the consistency of PCORM, which can be applied in general, does not make assumptions on the correct ordering or the position of the true MTC. In fact, the condition does not even require the exact correct ordering to be included into the POCRM. It suffices to include any ordering in the correct group.

The prior probabilities of orderings, as long as not 0 or 1, does not affect consistency. Nevertheless, they might affect the efficiency, and thus it is of interest to investigate efficient ways to assign prior probabilities. The examples presented in this paper assumes equal prior probabilities to all orderings included. An alternative approach is to assign priors based on \verb|n.consis|, it is left as a future work.

This paper focused on dual-agent trials. Nevertheless, partial ordering can be applied to any setting with uncertainty in orderings, such as the combination of drug and schedules, or more than two agents. Upon listing all orderings of the treatment regimens, the POCRM can be applied in the same way and the consistency conditions in this paper can be utilised to select ordering and improve efficiency. However, as the setting becomes complicated, it may be infeasible to list all orderings, which requires further investigation.

\section*{Acknowledgements}

This report is an independent research supported by the National Institute for Health Research (NIHR300576). The views expressed in this publication are those of the authors and not necessarily those of the NHS, the National Institute for Health Research or the Department of Health and Social Care (DHSC). PM also received funding from UK Medical Research Council (MC UU 00002/19). For the purpose of open access, the author has applied a Creative Commons Attribution (CC BY) licence to any Author Accepted Manuscript version arising. WC receives the Gates Cambridge Scholarship for her PhD.\vspace*{-8pt}

\clearpage
\bibliography{References}

\end{document}